\documentclass[11pt,a4paper]{article}

\usepackage{jheppub}
\usepackage{graphicx}
\usepackage{longtable}
\usepackage{amsmath}
\usepackage{amssymb}
\usepackage{hyperref}
\usepackage[noabbrev]{cleveref}
\usepackage{dcolumn}
\usepackage{bm}
\usepackage{color}

\newcommand{\be}{\begin{align}}
\newcommand{\ee}{\end{align}}
\newcommand{\mpl}{M_\mathrm{Pl}}
\newcommand{\Mp}{M_\mathrm{Pl}}

\newcommand{\dd}{\mathrm{d}}
\def\lsim{\mathrel{\mathstrut\smash{\ooalign{\raise2.5pt\hbox{$<$}\cr\lower2.5pt\hbox{$\sim$}}}}}
\def\gsim{\mathrel{\mathstrut\smash{\ooalign{\raise2.5pt\hbox{$>$}\cr\lower2.5pt\hbox{$\sim$}}}}}

\def\bea{\begin{align}}
\def\eea{\end{align}}

\newcommand{\nm}{{\mu\nu}}
\newcommand{\mn}{{\mu\nu}}
\newcommand{\ab}{{\alpha\beta}}
\newcommand{\oo}{\mathcal{O}}
\newcommand{\ooo}{{_{00}}}
\newcommand{\iij}{{_{ij}}}
\newcommand{\ii}{{_{ii}}}
\newcommand{\jj}{{_{jj}}}
\newcommand{\oi}{{_{0i}}}

\DeclareMathOperator{\Tr}{Tr}

\makeatletter
\DeclareRobustCommand{\rcite}[1]{%
  \rcite@aux#1,\@nil{#1}%
}
\def\rcite@aux#1,#2\@nil#3{%
  \if\relax#2\relax
    Ref.~\cite{#3}%
  \else
    Refs.~\cite{#3}%
  \fi
}
\makeatother

\begin{document}
\title{{Superfluids} and the Cosmological Constant Problem}
%\title{Supersolid Degravitation, or: How I Learned to Stop Worrying and Love Lorentz-Violating Massive Gravity}
%\title{Supersolid degravitation or: How I learned to stop worrying and love Lorentz-violating massive gravity}
\author{Justin Khoury,}
\author{Jeremy Sakstein, and}
\author{Adam R. Solomon}
\affiliation{Center for Particle Cosmology, Department of Physics and Astronomy,\\University of Pennsylvania, 209 S. 33rd St., Philadelphia, PA 19104, USA}
\emailAdd{jkhoury@physics.upenn.edu}
\emailAdd{sakstein@physics.upenn.edu}
\emailAdd{adamsol@physics.upenn.edu}

\abstract{We introduce a novel method to circumvent Weinberg's no-go theorem for self-tuning the cosmological vacuum energy: a Lorentz-violating finite-temperature superfluid can counter the effects of {an arbitrarily large} cosmological constant. Fluctuations of the superfluid result in the graviton acquiring a Lorentz-violating mass and we identify a unique class of theories that are pathology free, phenomenologically viable {and do not suffer from instantaneous modes.} This new and hitherto unidentified phase of massive gravity propagates the same degrees of freedom as general relativity with an additional Lorentz-violating scalar that is introduced by higher-derivative operators in a UV insensitive manner. The superfluid is therefore a consistent infrared modification of gravity. We demonstrate how the superfluid can degravitate a cosmological constant and discuss its phenomenology.}

\maketitle

\section{Introduction}

Quantum field theory and general relativity (GR) have been two cornerstones of theoretical physics throughout the last century, successfully predicting every empirical test that has been thrown at them. When applied to what is perhaps the simplest calculation imaginable, however, the energy of empty space, there is an ignominious puzzle: the cosmological constant problem~\cite{Weinberg:1988cp,Burgess:2013ara,Padilla:2015aaa}. The vacuum energy, which plays the role of the cosmological constant, is both UV sensitive and radiatively unstable. Integrating out heavy particles of mass $m$ results in contributions to the vacuum energy density of order $m^4$ at all loop orders in perturbation theory. The observation that the vacuum energy today is of order meV$^4$ (or possibly zero if dark energy is not a cosmological constant) implies that the contributions from all particles with masses larger than meV are being cancelled out by the cosmological constant above the cutoff. Given that the LHC has not observed any new physics at TeV scales, the cutoff for the standard model must be super-TeV.\footnote{The absence of any superparticles below the TeV scale~\cite{Agashe:2014kda} means that supersymmetry must be broken below this scale and therefore cannot alter this.} The cancellation is therefore exact to at least 60 decimal places (or upwards of 120 if the cutoff for the effective field theory is at the Planck mass) at each loop order in perturbation theory, a severe and repeated fine-tuning. This extreme UV sensitivity is the cosmological constant problem.

The cosmological constant problem is not easily mitigated since one would need to alter how particles with well-measured properties gravitate or how they contribute to the vacuum energy. Furthermore, any proposed solution must not introduce any additional fine-tuning. The complexity of the problem has inspired some truly ingenious solutions, including wormholes~\cite{Coleman:1988tj,Klebanov:1988eh}, which result in a probability distribution for the cosmological constant peaked at zero; mirror ghost universes~\cite{Linde:1988ws,Kaplan:2005rr}, which impose an energy parity symmetry that forbids a cosmological constant; large supersymmetric extra dimensions~\cite{Aghababaie:2003wz}, which have the visible sector living on a flat 4D brane that is a defect in a larger six-dimensional ``rugby ball'' shaped universe;\footnote{This is similar to how a cosmic string in four dimensions is a two-dimensional Ricci-flat defect.} composite metrics~\cite{Kimpton:2012rv}, where the metric depends on other fields in such a way that $\int\dd ^4 x\sqrt{-g}$ a topological invariant so that the vacuum energy is non-dynamical; and vacuum energy sequestering~\cite{Kaloper:2013zca,Kaloper:2014dqa,Kaloper:2016jsd}, where the classical part of the cosmological constant is promoted to a global variable whose variation ensures that both matter and graviton loops do not contribute to the vacuum energy (see \rcite{Kaloper:2015jra} for a local formulation).  

Another possibility is that there is no fine-tuning and the vacuum energy really is super-TeV, but its gravitational effects are mitigated by other fields. This is the case in relaxation models~\cite{Abbott:1984qf,Alberte:2016izw,Graham:2017hfr}, in which the value of the cosmological constant is selected dynamically, and self-tuning models. Self-tuning models posit new fields whose dynamics are such that the vacuum is Poincar\'{e}-invariant, which implies zero cosmological constant. Unfortunately, Weinberg's famous no-go theorem~\cite{Weinberg:1988cp} implies that any such attempt either fails because the fine-tuning remains, or the matter sector has a conformal symmetry (and therefore all masses are zero) which is clearly not the case in our universe. Of course, every no-go theorem starts from assumptions that one can try to break. In the case of Weinberg's theorem, the assumption most amenable to circumventing is that the vacuum is Poincar\'{e}-invariant. Indeed, examples that can successfully self-tune in this way, the \emph{fab four}~\cite{Charmousis:2011bf,Charmousis:2011ea,Copeland:2012qf} (and beyond fab four~\cite{Babichev:2015qma}), have been found. These include a new scalar that acquires a cosmological time-dependent expectation value and therefore breaks time translations. Recently, \rcite{Niedermann:2017cel} has provided further restrictions on Lorentz-invariant self-tuning models that utilize broken translations.

One can also {attempt to} construct theories that \emph{degravitate}~\cite{ArkaniHamed:2002fu,Dvali:2007kt}. In these theories, the cosmological constant is large compared with the observed vacuum energy today (as it is in self-tuning models), but gravity does not respond to it in the same manner as it does to other gravitating sources. In some sense, the equivalence principle between vacuum and non-vacuum energy is broken. In degravitation theories, the gravitational field equations contain a natural high-pass filter so that the contributions of long-wavelength modes (including the cosmological constant) are filtered out while short-wavelength modes are unaffected. {In the weak-field regime, any degravitating theory must reduce to a theory of massive gravity~\cite{Dvali:2007kt}.} The graviton mass is subject to stringent bounds~\cite{deRham:2016nuf}. The de Rham-Gabadaze-Tolley (dRGT) theory~\cite{deRham:2010kj,deRham:2014zqa}, the unique ghost-free Lorentz-invariant theory of a massive spin-2 particle, is unable to simultaneously degravitate a super-TeV cosmological constant and satisfy solar system tests of gravity~\cite{deRham:2010tw}. In particular, the Vainshtein radius, below which the van Dam-Veltman-Zakharov (vDVZ) discontinuity~\cite{vanDam:1970vg,Zakharov:1970cc} is resolved, would be too small to allow for general relativistic behavior to be recovered in the solar system.

In this paper, we introduce {a new model to circumvent Weinberg's no-go theorem:} \emph{degravitation with a finite-temperature superfluid}. Our model consists of a quartet of scalar fields $\Phi^A$ ($A=0,\,1,\,2,\,3$) that represent the internal degrees of freedom {(phonons)} of a cosmological self-gravitating superfluid. {The ground state of the superfluid is described by the scalar fields acquiring Poincar\'{e}-violating but $\mathrm{SO}(3)$-preserving vacuum expectation values:
\begin{equation}
\Phi^0 = \alpha t\,;\qquad \Phi^a = \beta x^a \,;\qquad a=1,\,2,\,3\,,
\end{equation}
where $\alpha$ and $\beta$ are constants. That is, the superfluid is isotropic. Moreover, the superfluid is such that its ground-state energy-momentum tensor takes the form of a cosmological constant, 
\begin{equation}
T_{\mu\nu} = - \Lambda_\mathrm{SF}(\alpha,\beta) \eta_{\mu\nu}\,.
\label{vacuumTmunu}
\end{equation}
By adjusting $\alpha$ and $\beta$, the ground state energy $\Lambda_\mathrm{SF}$ can cancel an arbitrarily large vacuum energy contribution from matter fields. This leads to a Minkowski solution regardless of the size of the vacuum energy. This cancellation is a result of the dynamics. That is, it is a property of the solution, for suitable initial conditions, and does not require that the parameters appearing in the Lagrangian be finely tuned against the cosmological constant. Of course, for this to qualify as a genuine solution to the cosmological constant problem, it remains to be shown that the Minkowski solution is a dynamical attractor for generic initial conditions. This will require an analysis of the cosmological evolution of the model, which we leave for future work. Nevertheless, as it stands, the theory does successfully achieve the goal set in Weinberg's theorem: for a fixed Lagrangian describing the superfluid, and in the presence of an arbitrarily large vacuum energy, one can find a ground-state configuration of the superfluid such that Minkowski space is a solution to the classical equations of motion.} 

In this sense, our theory is similar to self-tuning models---it evades the assumptions underlying Weinberg's no-go theorem because the $\Phi^A$'s spontaneously break Poincar\'{e} invariance. 
On the other hand, it is also akin to degravitation models because the scalars themselves are none other than the St\"{u}ckelberg fields for Lorentz-violating massive gravity. That is, fluctuations of the superfluid give rise to Lorentz-violating mass terms for the graviton. (The equivalence between Lorentz-violating massive gravity and self-gravitating media, {\it e.g.}, fluids, superfluids, and superfluids, is well-known and documented in the literature. See, for example, \rcite{Dubovsky:2004sg,Nicolis:2011cs,Ballesteros:2016gwc,Celoria:2017bbh,Celoria:2017idi,Celoria:2017hfd} and references therein.) This Lorentz violation also allows us to circumvent the no-go theorem of~\rcite{Niedermann:2017cel}. It is well-known that the vDVZ discontinuity can be absent in Lorentz-violating massive gravity theories~\cite{Rubakov:2004eb}, so that the problem which was debilitating for degravitation in the Lorentz-invariant case is not present in our model.

We set forth a set of theoretical and observational criteria for our model {to satisfactorily evade Weinberg's theorem.} While we will elaborate on these below, we summarize them in brief here:

\begin{enumerate}

\item The theory should degravitate dynamically, {\it i.e.}, it must not require parameters appearing in the Lagrangian to be finely tuned against the cosmological constant. 

\item It must be pathology free, in the sense that there should be no ghost, gradient, or tachyonic instabilities. Additionally, we require that the theory not permit instantaneously-propagating modes.

\item The model should be phenomenologically viable, {\it i.e.}, it should pass all current tests of gravitation. {In particular, to ensure that it satisfies the stringent bounds on the mass of tensor gravitons, we impose that the transverse, traceless modes (gravitational waves) are massless.}

\item The model should be UV insensitive, {\it i.e.}, the theory should be protected by a symmetry so that it is natural in the 't Hooft sense. Moreover, higher-derivative operators should not introduce degrees of freedom that are unsuppressed by the cutoff.

\end{enumerate}

We frame these criteria within the framework of Lorentz-violating massive gravity and are led to a unique and hitherto-unstudied region of parameter space protected by {time-dependent volume-preserving spatial diffeomorphisms,}
\begin{equation}
\partial_i \xi^i(t,\vec{x})=0\,. 
\label{timeindepVdiffintro}
\end{equation}
This is a Lorentz-violating form of the transverse diffeomorphism symmetry (TDiff) that is often studied as an alternative to GR and that underlies unimodular gravity.

{In terms of the St\"uckelberg fields, the symmetry~\eqref{timeindepVdiffintro} acts on the spatial scalars $\Phi^a$ as
\begin{equation}
\Phi^a\to \Psi^a(\Phi^0,\Phi^b)\,,\qquad\quad\det\left(\frac{\partial\Psi^a}{\partial\Phi^b}\right)=1\,.
\end{equation}
The invariant building blocks of this symmetry at leading order in the derivative expansion ({\it i.e.}, with one derivative per field) are
\begin{equation}
X\equiv g^\nm\partial_\mu\Phi^0\partial_\nu\Phi^0\,;\qquad Yb \equiv \sqrt{-\det \left(g^\nm\partial_\mu\Phi^A\partial_\nu\Phi^B\right)}\,.
\end{equation}
(In unitary gauge, where $\partial_\mu\Phi^a=\delta_\mu^a$, the latter reduces to the inverse metric determinant, $Yb = 1/\sqrt{-g}$.)
The most general action invariant under this symmetry, to leading order in the derivative expansion, is
\begin{equation}
\frac{\mathcal{L}}{\sqrt{-g}} =\frac{\mpl^2}{2} m^2U(X,Yb)\,.
\label{eq:theoryintro}
\end{equation}
This is a particular example of a finite-temperature superfluid action~\cite{Nicolis:2011cs}.} 

The $X$ dependence in \cref{eq:theoryintro} turns out to be essential to satisfy all of our criteria for successful degravitation. If $U$ were a function of $Yb$ only,
then the theory would be invariant under the (Lorentz-invariant) transverse diffeomorphism (TDiff) symmetry, $\partial_\mu \xi^\mu=0$~\cite{Henneaux:1989zc,Kuchar:1991xd,Padilla:2014yea}. As we will see, TDiff theories are UV sensitive: to lowest order in derivatives, they propagate the same number of degrees of freedom as GR, but higher-derivative corrections compatible with the symmetries resurrect a scalar mode that propagates at arbitrarily low energies, rather than being suppressed below the cutoff. Such a model is therefore not a sensible low-energy effective field theory.

In order for the superfluid theory~\eqref{eq:theoryintro} to have a ground-state energy-momentum tensor of the vacuum energy form~\eqref{vacuumTmunu}, it is necessary that
\begin{equation}
U_{X} = 0\,.
\end{equation}
This condition is akin to ghost condensation~\cite{ArkaniHamed:2003uy}. Indeed we will see that our model has much in common with the ghost condensate.
{In particular, there is a scalar mode with dispersion relation $\omega^2 = 0$ at lowest order in derivatives which becomes dynamical when higher-derivative terms compatible with the symmetries are included.} When these higher-derivative operators are included, we find similar behavior to ghost condensation, insofar as a new dynamical scalar degree of freedom appears and mixes with gravity. We find a similar Jeans-like instability at large distances (or at low momenta) that is stabilized by higher-order terms in the dispersion relation. This is not problematic, as strong Lorentz-violation effects imply that the instability is suppressed at early times on long distances. Our theory is therefore a consistent infrared modification of gravity. In a sense, our theory is to TDiff theories what the ghost condensate is to GR.

We will demonstrate how degravitation works explicitly in this theory using two simple working examples for $U(X,Yb)$.
We will present a full and comprehensive analysis of the cosmology of the superfluid in a forthcoming publication.

The paper is organized as follows. We start in \cref{sec:scalars} by examining a pair of simple scalar-field models which one might expect to violate Weinberg's no-go theorem, and discuss the flaws with each, leading us to consider breaking the Lorentz group to spatial rotations. In \cref{sec:LVMG} we give a brief introduction to self-gravitating media and their formulation as theories of Lorentz-violating massive gravity. We will present the St\"{u}ckelberg formalism which we use to construct $\mathrm{SO}(3)$-invariant combinations of four scalars that give rise to linearized Lorentz-violating massive gravity on Minkowski space when expanded in unitary gauge. These are the scalars we will use to search for healthy degravitating theories. In \cref{sec:criteria} we present our set of criteria for a theory to successfully degravitate. In \cref{sec:m20,sec:m10syms} we identify the symmetries that give rise to theories satisfying these criteria and show how they can be realized in the St\"{u}ckelberg formalism. This analysis reveals the unique symmetry we consider for a successful model, namely time-dependent volume-preserving spatial diffeomorphisms~\eqref{timeindepVdiffintro}. We then study the mixing with gravity once we include higher-derivative operators and discuss the phenomenology of our theory. In \cref{sec:degrav} we discuss degravitation in our theory and present two simple models satisfying all of our criteria. We summarize our results and identify future avenues of investigation in \cref{sec:conc}.

\section{Degravitation with multiple scalars}
\label{sec:scalars}

Our aim is to circumvent Weinberg's no-go theorem by allowing for some number of scalar fields to vary over space and time in vacuum. In order to ensure that the fields' stress-energy tensor cancels out that of a cosmological constant, they must depend on space-time in such a way that their stress-energy tensor is Lorentz-invariant. This implies that the fields must (at a minimum) be accompanied by derivatives. To leading order in derivatives, we can consider theories with one derivative per field. 

{In this section we walk through a sequence of unsuccessful examples, starting from the simplest case of a single scalar field with derivative interactions. The lessons learned along the way will lead us to the successful model described in the following section.} 

\subsection{One scalar}\label{sec:onescalar}

We begin by asking a simple question: can one find Minkowski solutions in the presence of a non-zero cosmological constant and a single scalar field? Generically this turns out to be impossible.\footnote{One exception to this are self-tuning models such as the ``fab four"~\cite{Charmousis:2011bf,Charmousis:2011ea} involving non-trivial curvature couplings, which are not the focus of the present work.} The most general Lagrangian with one derivative per field is of the $P(X)$ form, 
\begin{equation}
\frac{\mathcal{L}}{\sqrt{-g}}=\frac{\mpl^2}{2}\left[R - 2\Lambda + m^2P(X)\right]\,,
\end{equation}
where $X=(\partial\Phi)^2$. {The $P(X)$ Lagrangian describes a superfluid, with the scalar field excitations describing longitudinal phonons~\cite{Greiter:1989qb,Son:2002zn}.} To make contact with later discussions, we have chosen $\Phi$ to have units of length, so we include an arbitrary mass scale $m$ on dimensional grounds. The combined energy-momentum tensor for the cosmological constant and scalar field is
\begin{equation}
T_\mn = \Mp^2\left[ \frac12m^2\left(Pg_\mn - 2P_X\partial_\mu\Phi \partial_\nu\Phi\right)-\Lambda g_\mn \right]\,,
\end{equation}
where $P_X \equiv \dd P/\dd X$.\footnote{Throughout this paper, we denote differentiation by subscripts. Where a function depends on multiple variables, subscripts denote partial derivatives.}

In order to completely cancel the contribution of the cosmological constant without fine-tuning, we must find a solution $\Phi=\Phi(x^\mu)$ and $g_\nm=\eta_\nm$ such that $T_\mn=0$. We can split this into two requirements: $i)$ that the $\Phi$ part of $T_\mn$ be proportional to $\eta_\mn$, and $ii)$ that its coefficient be equal to $\Lambda$ without tuning any of the free parameters in $P(X)$. The first of these requires $P_X=0$, as it is plainly impossible to choose $\Phi$ such that $\partial_\mu\Phi\partial_\nu\Phi\propto\eta_\mn$. This is satisfied by ghost condensate models, in which $P_X=0$ but $X\neq0$ on the vacuum~\cite{ArkaniHamed:2003uy}. For example, we might have $P=X+\frac12 \lambda X^2$, such that $P_X=0$ is solved by $\Phi = \lambda^{-1/2} t$. We then violate our second condition, as $T_\mn$ only vanishes if $\lambda$ is carefully fine-tuned against $\Lambda$. In {general}, a given $P(X)$ clearly only degravitates one specific value of $\Lambda$.

\subsection{Four scalars: a simple example}\label{sec:fourscalars}

The problem with the above analysis was that we only had one scalar, which led to an inhomogeneous contribution $\sim \partial_\mu\Phi\partial_\nu\Phi$ to the energy-momentum tensor, precluding solutions where $T_\nm\propto\eta_\nm$ unless $P_X=0$. The obvious extension is to consider four scalars, so that the analogous term can itself be proportional to $\eta_\mn$.

As a simple example, let us imagine that the four scalars $\Phi^A$ have an internal {$\mathrm{SO}(1,3)$} Lorentz symmetry. The simplest invariant we can write down at lowest order in derivatives is
\begin{equation}
C \equiv g^\nm\partial_\mu\Phi^A\partial_\nu\Phi^B\eta_{AB}\,.
\end{equation}
To leading order in derivatives, the theory is given by an arbitrary function of $C$ coupled to gravity,
\begin{equation}\label{eq:Justin}
\frac{\mathcal{L}}{\sqrt{-g}}=\frac{\mpl^2}{2}\left[R - 2\Lambda + m^2U(C)\right]\,.
\end{equation}
Note that this theory has an $\mathrm{SO}(1,3)_\textrm{space-time}\times\mathrm{SO}(1,3)_\mathrm{internal}$ symmetry. The total energy-momentum tensor for the cosmological constant and the scalars is
\begin{equation}
T_\nm=\mpl^2\left[ m^2\left(\frac12Ug_\nm - U_C\eta_{AB}\partial_\mu\Phi^A\partial_\nu\Phi^B\right)-\Lambda g_\nm \right]\,.
\end{equation}

Even when this theory has a non-zero cosmological constant, it admits simple flat solutions which degravitate the cosmological constant given a particular vacuum solution for the scalars,
\begin{align}
g_\mn &= \eta_\mn\,; \\
\Phi^A &= \alpha x^A\,.
\label{fourscalarsoln}
\end{align}
The integration constant $\alpha$ is related to the cosmological constant that it removes by
\begin{equation}
\frac12U - \alpha^2U_C = \frac{\Lambda}{m^2}\,,
\end{equation}
where $U$ and $U_C$ are evaluated on $C=4\alpha^2$. The solution~\eqref{fourscalarsoln} breaks the $\mathrm{SO}(1,3)_\textrm{space-time}\times\mathrm{SO}(1,3)_\mathrm{internal}$ to a diagonal $\mathrm{SO}(1,3)$ subgroup.

For example, for the simplest choice, $U(C)=\pm C$, we can degravitate a positive (negative) cosmological constant with $\alpha=\sqrt{|\Lambda|}/m$. Crucially, degravitation occurs here because of the integration constant, without having to tune the free parameters (if there are any) in $U(C)$ against $\Lambda$, allowing us to screen out an arbitrary $\Lambda$.\footnote{Potentially for a particular sign of $\Lambda$, as the simple example $U(C)=\pm C$ showed.} This consideration will be crucial throughout our analysis.

This model turns out to be too simple, as it suffers from a ghost instability. We can see this by phrasing this theory in terms of massive gravity. To do that, consider perturbations of the form
\begin{equation}
\Phi^A = \alpha(x^A+\pi^A)\,.
\end{equation}
We can change coordinates to unitary gauge, where all of the the perturbations are contained in the metric, by sending
\begin{equation}
x^\mu \to \tilde{x}^\mu = x^\mu + \pi^\mu\,.
\end{equation}
Let us focus on scalar perturbations with $\pi_\mu\equiv\partial_\mu\pi$, where we define $\partial_\mu$ to be the partial derivative operator with respect to $\tilde x^\mu$. To be fully general we should decompose $\pi_\mu$ as $A_\mu + \partial_\mu\pi$, where $\partial_\mu A^\mu=0$, but including the vector will only modify our final results by a healthy Maxwell term for $A^\mu$, so we can safely ignore it~\cite{deRham:2014zqa}. In this gauge, the flat metric takes the form
\begin{align}
g_\mn &= \eta_\ab \frac{\partial x^\alpha}{\partial\tilde x^\mu} \frac{\partial x^\beta}{\partial\tilde x^\nu} \nonumber \\
&= \eta_\mn -2\Pi_\mn + \Pi^2_\mn\,,
\end{align}
where $\Pi_\mn\equiv\partial_\mu\partial_\nu\pi$, and the indices are raised and lowered with $\eta_\nm$. We can write this metric in the form
\begin{equation}
g_\mn = \eta_\mn + h_\mn + \frac14h_{\mu\alpha}h^\alpha{}_\nu\,,
\end{equation}
where we identify $h_\mn=-2\Pi_\mn$. We can then write $C$ in terms of $h_\mn$,
\begin{equation}
C = \alpha^2\eta^\mn g_\mn = \alpha^2\left(4-[h]+\frac14[h^2]\right)\,,
\end{equation}
where we have defined $[h]\equiv \eta_\mn h^\mn$ and $[h^2]\equiv h_\mn h^\mn$. In this gauge, $U(C)$ amounts to a mass term for $h_\mn$ at quadratic order,
\begin{equation}
\mathcal{L}_\mathrm{mass} = \alpha^2\left(U_C[h^2] + 2\alpha^2U_{CC}[h]^2\right)\,.
\end{equation}
It is well-known that such a mass term propagates a ghost unless $[h^2]$ and $[h]^2$ arrange themselves into the Fierz-Pauli combination $[h^2]-[h]^2$. This requires $U(C)$ to satisfy the condition $U_C + 2\alpha^2U_{CC}=0$, which is generically not the case.\footnote{This condition is satisfied by the singular function $U(C)\sim C^{-1}$. Regardless, even if we were to choose $F(C)$ to obtain the Fierz-Pauli tuning, the ghost generically reappears at the non-linear level unless we choose the special dRGT non-linear mass term, which cannot be written {solely} in terms of $C$~\cite{deRham:2010kj,Hinterbichler:2011tt,deRham:2014zqa}.}

This instability is tied to the choice of $\eta_{AB}$ for the field-space metric. This is particularly clear in the simple case $U=-C$, so that the action is
\begin{equation}
\frac{\mathcal{L}}{\sqrt{-g}}=\frac{\mpl^2}{2}\left[R - 2\Lambda +m^2\Big(\partial_\mu\Phi^0\partial^\mu\Phi^0-\delta_{ij}\partial_\mu\Phi^i\partial^\mu\Phi^j\Big)\right]\,.
\end{equation}
Clearly $\Phi^0$ has a wrong-sign kinetic term and is therefore a ghost. This is a direct result of choosing $\eta_{AB}$ as the internal metric for the $\Phi^A$: it is impossible for all four scalars to have the correct-sign kinetic terms. Recall that we made this choice in order to facilitate degravitation, as it allows $T_\nm\propto \eta_\nm$ after imposing the condition $\partial_\mu\Phi^A=\alpha \delta^A_\mu$. This simple model is unstable for the same reason that it is able to degravitate in the first place.

This then raises the question of whether one can find other combinations of kinetic terms for the fields $\Phi^A$ such that we can cancel the cosmological constant from the energy-momentum tensor without inducing instabilities; this is one of the central questions of this paper, and we will find that the answer is indeed yes.

{\subsection{Four scalars and Lorentz-invariant massive gravity}}

We begin by making two important observations. First, it was not necessary to break the $\mathrm{SO}(1,3)_\textrm{space-time}\times\mathrm{SO}(1,3)_\mathrm{internal}$ symmetry to the full diagonal $\mathrm{SO}(1,3)$. We could just as well have preserved the diagonal $\mathrm{SO}(3)$ while breaking boosts. In terms of the $\Phi^A$, this means treating $\Phi^0$ separately from the spatial $\Phi^a$, where $a=1,\,2,\,3$, while contracting spatial indices with $\delta_{ab}$ and $\epsilon_{abc}$.\footnote{We could, of course, break the $\mathrm{SO}(3)$ further, but this would pose a problem for homogeneous and isotropic cosmological solutions.}

The second observation, as we have already alluded to, is that the theory~\eqref{eq:Justin} is in fact a (non-linear) massive gravity theory, where the reference metric is flat and we are working in the St\"{u}ckelberg language. To see this, note that one could obtain a massive gravity theory by augmenting the Einstein-Hilbert action with a potential for the metric~$g_\nm$ of the form
\begin{equation} \label{eq:simplemgaction}
\frac{\mathcal{L}}{\sqrt{-g}}=\frac{\mpl^2}{2}\left[R - 2\Lambda + m^2U(g^\mn \eta_\mn)\right]\,.
\end{equation}
This explicitly breaks diffeomorphism invariance due to the presence of $\eta_\mn$, which is necessary to construct non-trivial, non-derivative interactions for $g_\mn$. However, diffeomorphism invariance is a gauge symmetry, that is, a redundancy in description, and can always be restored by adding redundant variables. This is known as the St\"uckelberg trick, which can be done by performing a gauge transformation and promoting the transformation variables to fields with their own transformation properties under the gauge symmetry. In this case, we can achieve this by introducing ``coordinates" $\Phi^A$ and replacing $\eta_\mn\rightarrow \eta_{AB}\partial_\mu\Phi^A\partial_\nu\Phi^B$. As long as the $\Phi^A$ fields transform as space-time scalars, the action is diffeomorphism-invariant, and indeed is precisely the action \eqref{eq:Justin}. The action \eqref{eq:simplemgaction} can be recovered by choosing the so-called unitary gauge, in which $\Phi^A=x^A$, which can always be reached via an appropriate coordinate transformation as long as $\partial_\mu\Phi^A$ is non-singular.

This observation explains why one of our fields is a ghost. Massive gravity theories beyond the linear level are generically plagued by a ghostly degree of freedom, the Boulware-Deser ghost~\cite{Boulware:1973my}. The unique Lorentz-invariant theory of massive gravity which avoids this ghost is that of de Rham, Gabadadze, and Tolley (dRGT)~\cite{deRham:2010kj,deRham:2014zqa}. The dRGT theory has a rather more intricate structure than the simple example \eqref{eq:simplemgaction}.

One could search for ghost-free degravitating solutions within dRGT massive gravity, as was investigated in \rcite{deRham:2010tw}. While such solutions exist in dRGT, they cannot be reconciled with solar systems tests of gravity unless the degravitated cosmological constant is less than order meV$^4$. In particular, one cannot find a region of parameter space that simultaneously degravitates the large cosmological constant predicted by the standard model while also possessing a sufficiently efficient Vainshtein mechanism to suppress the ``fifth force" in the solar system.

Combining these two observations, we are led to an alternative and hitherto-unexplored possibility: to degravitate a large cosmological constant by breaking Lorentz invariance in the gravity sector while retaining global $\mathrm{SO}(3)$ invariance. In this approach, one is studying a Lorentz-violating theory of a massive graviton on a Minkowski background (given by the degravitated solution). It is well-known that the Boulware-Deser ghost can be exorcised in Lorentz-violating theories, and that the van Dam-Veltman-Zakharov (vDVZ) discontinuity is absent~\cite{Rubakov:2004eb,Dubovsky:2004sg,Comelli:2013paa}. In particular, one does not need the Vainshtein mechanism~\cite{Vainshtein:1972sx}. One might therefore hope that the obstructions to degravitation encountered in dRGT models can be circumvented in Lorentz-violating massive gravity.\\

\section{Self-gravitating media and Lorentz-violating massive gravity}
\label{sec:LVMG}

Breaking Lorentz invariance increases the number of interactions that one can write down. It is convenient to classify the most general action into ``phases" depending on which, if any, subgroup of the full diffeomorphism group one wishes to retain. In our case we choose to preserve $\mathrm{SO}(3)$ rotational symmetry at the least. This was extensively studied by Dubovsky~\cite{Dubovsky:2004sg}, who classified various phases within boost-violating massive gravity and examined the stability and UV sensitivity of each. Our strategy will be to specify a set of criteria required for degravitation and search for phases which meet them. {Ultimately, these criteria will lead us to a unique phase, as well as a unique symmetry which imposes it.} We will then use this symmetry to construct models that can successfully degravitate. The degravitating solutions are stable in the sense that any ghost, gradient, and tachyonic instabilities are absent. The theory is phenomenologically-viable and is UV insensitive, in a manner that we make precise below.

We begin in this section by reviewing the properties of theories of four scalars with internal $\mathrm{SO}(3)$ symmetry, focusing on their interpretations (depending on the choice of coordinates) as either self-gravitating mechanical continua (supersolids) or, equivalently, Lorentz-violating mass terms for the graviton.

Consider a quartet of derivatively-coupled scalar fields $\Phi^A(x^\mu)$. We can think of these as the comoving coordinates for a self-gravitating medium. Additionally, they encode the degrees of freedom required to describe this medium at low energy {and zero temperature}. As long as we choose to maintain an internal SO(3) rotational symmetry for these scalars, the medium that they describe can be thought of as an isotropic supersolid. As we will soon see, thanks to diffeomorphism invariance, when coupling to gravity we can always work in unitary gauge where the $\Phi^A$'s are unperturbed. In this case the self-interactions of the medium take on the form of non-derivative interactions ({\it i.e.}, mass terms) for the metric~\cite{Ballesteros:2016gwc}. Equivalently, we could go in the other direction: starting with a {general (Lorentz-violating) theory of massive gravity, we can restore diffeomorphism invariance by introducing St\"uckelberg fields, which are precisely the four scalar fields $\Phi^A$.} 

In order to construct the low-energy effective action for this supersolid, we focus on the leading-order terms in the derivative expansion. These can be completely characterized by the matrix
\begin{equation}
C^{AB}=g^{\mu\nu}\partial_\mu\Phi^A\partial_\nu\Phi^B\,.
\label{eq:CAB}
\end{equation}
This is nothing other than the inverse metric on the manifold defined by the coordinates $\Phi^A$, which can be induced by pulling back the metric onto that manifold.\footnote{Equivalently, we could start by writing
\[ \dd s^2=g_\mn\dd x^\mu \dd x^\nu = g_\mn\frac{\partial x^\mu}{\partial\Phi^A}\frac{\partial x^\nu}{\partial\Phi^B}\dd\Phi^A\dd\Phi^B\equiv C_{AB}\dd\Phi^A\dd\Phi^B,\]
which implies $C_{AB} = g_\mn\frac{\partial x^\mu}{\partial\Phi^A}\frac{\partial x^\nu}{\partial\Phi^B}$. Inverting this, \cref{eq:CAB} follows.} Because each of the $\Phi^A$ transforms as a scalar, $C^{AB}$ is also a scalar under space-time diffeomorphisms. 

If the internal symmetry is Lorentz, we can contract $C^{AB}$ with $\eta_{AB}$ and $\epsilon_{ABCD}$ as we please to construct a Lagrangian. {The result is equivalent to a Lorentz-invariant theory of massive gravity.} This is easy to see in unitary gauge in which $\partial_\mu\Phi^A=\delta^A_\mu$, so that $C^{AB}=g^{AB}$, and the Lorentz-invariant contractions of $C^{AB}$ amount to non-derivative interactions for~$g_\mn$. Almost all of the Lorentz-invariant Lagrangians one can write down, with the exception of the special case of dRGT massive gravity, will propagate the Boulware-Deser ghost and are therefore not healthy physical theories. {Furthermore, as we discussed in the previous section, even though dRGT massive gravity is classically healthy, it is unable to degravitate a large vacuum energy while passing solar system tests of gravity.}

Hence we will consider breaking the boost part of the internal Lorentz symmetry while leaving spatial rotations intact. In order to do this we further decompose $C^{AB}$ into components that are $\mathrm{SO}(3)$-invariant. We will write down a complete basis of such operators, largely following~\rcite{Ballesteros:2016gwc}.

We begin by distinguishing between the field $\Phi^0$, which corresponds to a time-like fluid coordinate, and the spatial fields $\Phi^a$, $a=1,\,2,\,3$, which retain an internal $\mathrm{SO}(3)$ symmetry. The spatial components of $C^{AB}$ define a spatial metric,
\begin{equation}\label{eq:Bdef}
B^{ab}\equiv C^{ab}\,.
\end{equation}
This metric is assumed positive-definite, so that it can be interpreted as a Euclidean metric on~$\mathbb{R}^3$. In particular, one can consider the coordinates $\Phi^a$ to span a three-dimensional manifold $\mathcal{F}$ defined by
\begin{equation}
\dd s^2_{\mathcal{F}}=g_\nm\frac{\partial x^\mu}{\partial\Phi^a}\frac{\partial x^\nu }{\partial \Phi^b}\dd\Phi^a\dd\Phi^b = B_{ab}\dd\Phi^a\dd\Phi^b \,,
\end{equation}
so that $B_{ab}$ is the induced metric on $\mathcal{F}$.

We can construct $\mathrm{SO}(3)$ invariants from $B_{ab}$ via contractions with $\delta_{ab}$ and $\epsilon_{abc}$. Since $B$ is a $3\times3$ symmetric matrix, it can be fully determined by its three eigenvalues, or equivalently, by its first three traces,
\begin{equation}
\tau_n\equiv\Tr(B^n)\,;\qquad n\leq3 \,.
\end{equation}
We will find it convenient later to work with the determinant
\begin{equation}
b\equiv\sqrt{\det{B}} = \sqrt{\frac{\tau_1^2-3\tau_1\tau_2+2\tau_3}{6}}\,,
\end{equation}
so we can fully characterize $B^{ab}$ by the invariants $\tau_1$, $\tau_2$, and $b$. The choice of invariants is often motivated by the symmetry-breaking patterns one wishes to study.

Two further $\mathrm{SO}(3)$-invariant quantities can be defined using the time-like field $\Phi^0$,
\begin{align}
X&\equiv g^\nm\partial_\mu\Phi^0\partial_\nu\Phi^0\,; \\
Y&\equiv u^\mu\partial_\mu\Phi^0\,,
\end{align}
where we have defined the four-velocity $u^\mu$ by\footnote{Note that $\epsilon^{\mu\nu\rho\sigma}$ here is the Levi-Civita symbol, not tensor.}
\begin{equation}\label{eq:udef}
u^\mu\equiv-\frac{\varepsilon^{\mu\nu\rho\sigma}\varepsilon_{abc}}{6\sqrt{-g} b}\partial_\nu\Phi^a\partial_\rho\Phi^b\partial_\sigma\Phi^c\,,
\end{equation}
with the normalization chosen such that $g_{\mu\nu}u^\mu u^\nu=-1$ and $u^\mu\partial_\mu\Phi^a=0$.\footnote{Later we will focus on the combination $Yb$, which has some special properties. By contracting epsilons, we can write this in terms of the determinant of $\partial_\mu\Phi^A$,
\[ Yb = \frac{\det{\partial\Phi}}{\sqrt{-g}}\,, \]
or manifestly as a space-time scalar,
\[ Yb = \sqrt{-\det C^{AB}}\,. \]
In unitary gauge, we have $\partial_\mu\Phi^a=\delta_\mu^a$ and therefore this is the determinant of the square-root matrix $\sqrt{g^{-1}\eta}$,
\[ Yb = \det\sqrt{g^{-1}\eta}\,. \]
This square-root matrix is reminiscent of dRGT massive gravity, and indeed such a term is one of the allowed ghost-free potential terms, although it is non-dynamical, $\mathcal{L} \sim \sqrt{-g}\sqrt{g^{-1}\eta}=\sqrt{-\eta}$. If $\eta_\mn$ is promoted to a dynamical metric $f_\mn$, as in bigravity \cite{Hassan:2011zd,Solomon:2015hja,Schmidt-May:2015vnx}, then this is a cosmological constant term for that metric.} 
Note that this is the unique vector with these properties that one can construct from the St\"{u}ckelberg fields. 
One can also construct a second four-velocity
\begin{equation}\label{eq:vdef}
v_\mu=\frac{\partial_\mu\Phi^0}{\sqrt{-X}}\,,
\end{equation}
which also has the property $g_{\mu\nu}v^\mu v^\nu=-1$. One can equivalently view $v_\mu$ as the normal to time-like hypersurfaces defined by $\Phi^0=$ constant. Both interpretations will be useful in what follows.

Finally, one can define a mixed temporal-spatial tensor,
\begin{equation}
Z^{ab}\equiv C^{0a}C^{0b}\,,
\end{equation}
which transforms as a tensor under $\mathrm{SO}(3)$. From this we can construct scalars by contracting powers of $B^{ab}$ with $Z^{ab}$,
\begin{equation}
y_n \equiv \Tr(B^n\cdot Z)\,,
\end{equation}
of which $y_0$, $y_1$, $y_2$, $y_3$, and $y_4$ are independent. This completes a basis of nine independent SO(3)-invariant operators,
\begin{equation}
\mathcal{O} = \{X,\,Y,\,b,\,\tau_1,\,\tau_2,\,y_0,\,y_1,\,y_2,\,y_3\}\,.
\end{equation}

Therefore, the most general action that is invariant under space-time diffeomorphisms and internal $\mathrm{SO}(3)$ symmetry is
\begin{equation}\label{eq:LVMGscalars}
S=\frac{\mpl^2}{2}\int\dd^4x\sqrt{-g}\left[R-2\Lambda+m^2U(X,Y,b,\tau_n,y_n)\right]\,.
\end{equation}
Diffeomorphism invariance is spontaneously broken by the scalar fields acquiring $\mathrm{SO}(3)$-invariant expectation values given by
\begin{equation}\label{eq:phiVEV}
\Phi^0=\alpha t\,;\qquad \Phi^a=\beta x^a\,.
\end{equation}
{Note that the symmetry-breaking patterns will differ depending on which of $X$, $Y$, $b$, $\tau_n$, and $y_n$ are included in the action. One is always guaranteed an $\mathrm{SO}(3)$-invariant vacuum by construction but further subgroups of the diffeomorphism group may remain.}

We are now in a position to connect this description of the $\Phi^A$ supersolid to (Lorentz-violating) massive gravity, {focusing specifically on linearized massive gravity. 
We assume that the equations of motion admit a flat space-time background solution $g_\nm=\eta_\nm$ for the supersolid ground state configuration~\eqref{eq:phiVEV}.}
We expand the metric and the scalars around their background solutions,
\begin{align}\label{eq:phiVEV2}
\Phi^0&=\alpha t+\pi^0\,; \nonumber \\
\Phi^a&=\beta x^a+\pi^a\,;\nonumber \\
g_\nm&=\eta_\nm+h_\nm\,,
\end{align}
where the $\pi^A$ can be viewed as the Goldstone fields of the spontaneously-broken Lorentz symmetry. The fields $h_\mn$ and $\pi^A$ are not all independent degrees of freedom because diffeomorphism invariance provides a redundancy; in fact, we can gauge away $\pi^A$ entirely with a change of coordinates. In this unitary gauge, the quadratic action amounts to deforming the (linearized) Einstein-Hilbert action with a potential for $h_\mn$. The most general mass term one can obtain from this procedure, retaining the $\mathrm{SO}(3)$ symmetry, can be written in the form~\cite{Rubakov:2004eb,Dubovsky:2004sg}
\begin{equation}\label{eq:LVMGaction}
\mathcal{L}_\mathrm{mass}=\frac{\mpl^2 }{2}\left[m_0^2h_{00}h_{00}+2m_1^2h\oi h\oi-m_2^2h\iij h\iij+m_3^2h\ii h\jj-2m_4^2h\ooo h\ii\right]\,.
\end{equation}

We will frequently work with the action \eqref{eq:LVMGaction} for various choices of $m_i^2$, or ``phases." The number of propagating degrees of freedom depends on which {(if any)} of the $m_i^2$ are non-zero, and these may or may not be healthy (in the sense of not possessing ghost, gradient, or tachyonic instabilities). Moreover, while this action breaks the full diffeomorphism invariance of the Einstein-Hilbert action, it may retain invariance under certain subgroups for different choices of $m_i^2$. Indeed, these two considerations will often go hand-in-hand: for reasons of stability and consistency with observations we may want to set certain of the $m_i^2$'s to zero, and the natural way to do this is by imposing a symmetry.

Any given function $U(X,Y,b,\tau_n,y_n)$ will lead to particular values of {the masses}, given by combinations of $U$ and its derivatives evaluated on the background \eqref{eq:phiVEV}. In particular, we have
\begin{align}\label{eq:fluidperts}
X&=-\alpha^2\left(1+h\ooo+h\ooo h\ooo-h\oi h\oi\right)\,;\nonumber\\
Y&=\alpha\left(1+\frac{1}{2}h\ooo +\frac{3}{8}h\ooo h\ooo\right)\,;\nonumber\\
b&=\beta^3\left[1-\frac{1}{2}h\ii+\frac{1}{8}\left(h\ii h\ii-4h\oi h\oi +2h\iij h\iij\right)\right]\,.
\end{align}
It will not be necessary to specify expansions for $\tau_n$ or $y_n$ for what follows.

\section{Criteria for degravitation}
\label{sec:criteria}

Having introduced the effective field theory for supersolids and its interpretation in terms of Lorentz-violating massive gravity, we now list our theoretical and observational criteria for successful degravitation:

\begin{enumerate}
\item {\bf Existence of a Minkowski (degravitating) solution}: The equations of motion must have a solution with $g_\nm=\eta_\nm$ for arbitrary $\Lambda$ (or at least for a wide range of~$\Lambda$), with the scalars taking the form~\eqref{eq:phiVEV}. Whether or not degravitation can occur is a property of the specific model, {\it i.e.}, the choice of $U$. \label{item:degrav}

\item {\bf Absence of fine-tuning}: In addition to simply possessing Minkowski solutions, degravitation must be a property of the solution \eqref{eq:phiVEV}, through the arbitrary integration constants $\alpha$ and $\beta$, rather than a consequence of tuning parameters in the Lagrangian against a specific choice of $\Lambda$. Tuning the free parameters would simply shift the radiative instability of the cosmological constant into the scalar sector, which is not a satisfactory solution of the cosmological constant problem.  \label{item:finetune}

\item {\bf Absence of massive tensors}: The mass of tensor modes is given by $m_2$. Recent bounds from the observation of merging black holes and neutron stars by the LIGO/Virgo collaboration~\cite{Abbott:2016blz,TheLIGOScientific:2016src,TheLIGOScientific:2017qsa,GBM:2017lvd,Sakstein:2017xjx,Baker:2017hug,Creminelli:2017sry,Ezquiaga:2017ekz,Langlois:2017dyl,Crisostomi:2017lbg} indicate that the mass of gravitons in the dispersion relation $\omega^2=p^2+m^2$ should be smaller than $10^{-22}$ eV.\footnote{Other stronger bounds may not apply due to their being linked with the decoupling limit of DGP braneworld or dRGT models~\cite{deRham:2016nuf}.} We generically expect that $m_i^2\sim\Lambda$ for any degravitating solution, barring fine-tuning.\footnote{This expectation arises on dimensional grounds, since the stress-energy tensor for $\Lambda$ and the scalar fields is, heuristically, of order $\Mp^2(\Lambda + m_i^2)$, and on the degravitating solution we require this combination to vanish. This expectation is borne out by worked examples, such as those in~\cref{sec:degrav}.} For this reason, we will impose that tensors are strictly massless. This will turn out to be rather restrictive, as it requires a symmetry that imposes $m_2=0$. \label{item:GWmass}

\item {\bf Absence of pathologies}: We demand that the theory not be pathological, in the sense that there are no tachyonic, ghost, or gradient instabilities, as well as any infinitely-strongly-coupled modes or instantaneous modes.\footnote{By instantaneous modes, we refer to modes that contain $\omega$-independent terms in their propagator (and therefore have time-independent contributions to their Green's functions) whose effects on physical observables do not vanish for time-dependent sources. Such modes transfer information instantaneously. While they do not violate causality~\cite{Dubovsky:2007ac}, as we have already violated Lorentz invariance, they are nonetheless disconcerting since they can give rise to ``bumpy black holes" (where one can probe inside the horizon)~\cite{Dubovsky:2007zi} and action-at-a-distance-type phenomena. Examples of such phenomena in Lorentz-violating electrodynamics have been studied in \rcite{Gabadadze:2004iv,Dvali:2005nt}.} This is a property of both the specific phase of massive gravity under consideration, {\it i.e.}, the number of propagating degrees of freedom, as well as the specific choice for the function $U$, as some phases may only be stable in specific regions of parameter space. \label{item:stability}

\item {\bf UV insensitivity}: We require that our theory be UV insensitive in the sense of Dubovsky~\cite{Dubovsky:2004sg}. A generic metric theory of gravity can propagate at most six degrees of freedom, but the imposition of symmetries that are subsets of the diffeomorphism group can reduce these. Higher-derivative corrections consistent with the symmetries may reintroduce some of these modes. We define a UV insensitive theory as one where such modes either are not reintroduced, or are reintroduced but with propagation suppressed by the cutoff $\mathcal{M}$. In practice this means that either their mass is of order $\mathcal{M}$ or their sound speed $\sim \omega^2/p^2$ is suppressed by $\mathcal{M}$. {In contrast, UV sensitive theories require fine-tuning at each order in perturbation theory to prevent these modes from being resurrected.} \label{item:uvsens}

\end{enumerate}

We can immediately draw some general conclusions about allowed models from criteria~\ref{item:finetune} and \ref{item:GWmass}. Criterion~\ref{item:finetune} requires that both integration constants, $\alpha$ and $\beta$, be present. This is because there are two conditions for degravitation, the requirement that $T_\nm\propto\eta_\nm$ and the requirement that the proportionality constant be $-\Lambda\mpl^2$, cf. the discussion in \cref{sec:fourscalars}. The absence of one or both integration constants would imply that fine-tuning is required in order to match one or both of these conditions.

Since we insist on an internal SO(3) symmetry, the three spatial scalars $\Phi^a$ can only provide one of these integration constants, and only $\Phi^0$ is left to provide the other. We therefore see that we need all four scalars $\Phi^A$ to be present in the action. In the language of self-gravitating media this implies that media with \emph{reduced internal dimensionality} (such as solids) are not viable, as these only have either the one time-like scalar or the three space-like scalars.

Criterion~\ref{item:GWmass} necessitates the use of a symmetry that imposes\footnote{We require that $m_2 = 0$ be the result of a symmetry, rather than of a particular parameter choice in the Lagrangian, in order to avoid fine-tuning.}
\begin{equation}
m_2=0\,.
\end{equation} 
The unique symmetry that {sets $m_2 = 0$, and only $m_2 = 0$,} is invariance under time-independent volume-preserving spatial diffeomorphisms. We will therefore begin by examining the phase $m_2=0$ in the next section and find that there is a generic instability in the scalar sector unless we further impose $m_1=0$. In the subsequent section, we will identify several additional symmetries that set $m_1=0$, and will ultimately find that only one leads to models that are both stable and UV insensitive. 

\section{Time-independent volume-preserving spatial diffeomorphisms}
\label{sec:m20}

We have established that criteria~\ref{item:finetune} and \ref{item:GWmass} require that we utilize the four scalars $\Phi^A$ and that the tensor mass $m_2=0$. 
The unique symmetry that {forces $m_2$, and only $m_2$, to vanish} is time-independent volume-preserving spatial diffeomorphism invariance (V$_\mathrm{s}$Diff)~\cite{Dubovsky:2004sg},
\begin{equation}
x^i\rightarrow \tilde{x}^i(x^j)\quad\textrm{with}\quad\det\left(\frac{\partial\tilde{x}^i}{\partial x^j}\right)=1 \,.
\end{equation}
At the linear level this corresponds to invariance under linear spatial diffeomorphisms $x^i\rightarrow x^i + \xi^i(x^j)$ with $\partial_i\xi^i=0$.
Equivalently, in terms of the St\"{u}ckelberg fields,
\begin{equation}
\Phi^a\rightarrow \Psi^a(\Phi^b)\quad\textrm{with}\quad\det\left(\frac{\partial\Psi^a}{\partial\Phi ^b}\right)=1\,.
\end{equation}
The invariants that preserve this symmetry are $X$, $Y$, and $b$. {Hence the most general action invariant under the assumed symmetries is
\begin{equation}
\frac{\mathcal{L}}{\sqrt{-g}} =\frac{\mpl^2}{2} m^2U(X,Y,b)\,.
\label{finiteTsuperfluid}
\end{equation}
}This is nothing other than the low-energy effective theory describing a finite-temperature relativistic superfluid~\cite{Nicolis:2011cs}.

\subsection{Stability properties}

To begin, we review the decoupling limit of this phase, {\it i.e.}, the Goldstone sector in the absence of any mixing with gravity. This was studied by Dubovsky~\cite{Dubovsky:2004sg}, who found that some amount of tuning is required to make the theory pathology-free. In particular, while the vectors do not propagate, one of the two scalar modes is generically a ghost, which has to be exorcised from the spectrum by tuning parameters so that this mode does not propagate.\footnote{There is no symmetry that can enforce this tuning.} The tuned theory that is healthy in the decoupling limit therefore propagates precisely one scalar mode. Because of this required tuning, the full theory including the mixing with gravity has, to our knowledge, never been studied in detail.

Let us include gravity and linearize around the flat solution, cf. \cref{eq:phiVEV2}. As discussed above, we can always choose the unitary gauge in which $\pi^A=0$ so that the linearized action is of the form \eqref{eq:LVMGaction} with $m_2=0$, in addition to the Einstein-Hilbert term. We then perform a scalar-vector-tensor decomposition of the metric,
\begin{align}
h\ooo &= \psi\,; \nonumber\\
h\oi & = u_i+\partial_iv\,;\nonumber\\
h\iij &= \chi\iij + \frac{1}{2}\partial_{(i}s_{j)}+\partial_i\partial_j\sigma+\delta\iij\tau\,,
\label{eq:metricexp}
\end{align}
where $u_i$ and $s_i$ are transverse vectors, and $\chi\iij $ is a transverse, traceless tensor. We will analyze the dynamics in this gauge, setting $m_2=0$ in \cref{eq:LVMGaction}. The tensor action is simply
\begin{equation}
\mathcal{L}_\mathrm{tensor} = \frac{\mpl^2}{4}\chi\iij\Box\chi\iij \,.
\end{equation}
{Thus, as desired, tensor modes propagate as in general relativity.} In the vector sector, the field $u_i$ is auxiliary, and after its elimination one is left with the action
\begin{equation}
\mathcal{L}_\mathrm{vector}=-\frac{1}{2}\partial_0s^\mathrm{c}_i\partial_0s^\mathrm{c}_i\,,
\end{equation}
where $s^\mathrm{c}_i$ is the canonically normalized form of the vector $s_i$~\cite{Rubakov:2004eb}. The dispersion relation is then $\omega^2=0$ and the vector modes do not propagate, just as in the decoupling limit. Finally, the scalar Lagrangian is\footnote{Note that this differs from several other forms appearing in the literature~\cite{Rubakov:2004eb,Dubovsky:2004sg,Dubovsky:2005dw} because in those cases the Einstein-Hilbert Lagrangian was taken to be $\mpl^2\sqrt{-g} R$ and, in some cases~\cite{Dubovsky:2005dw}, the coefficient of the mass Lagrangian was taken to be $\mpl^2/4$ rather than $\mpl^2/2$. We choose to use the canonical forms here.}
\begin{align}\label{eq:scalar1}
\mathcal{L}_\mathrm{scalar} &= \frac{\mpl^2}{2}\Big[\left(\psi-2\partial_0v+\partial_0^2\sigma\right)\partial_i^2\tau+\frac{3}{2}\tau\partial_0^2\tau-\frac{1}{2}\tau\partial_i^2\tau+m_0^2\psi^2+2m_1^2(\partial_iv)^2 \nonumber\\
&\hphantom{{}=\frac{\mpl^2}{2}\Big[} +m_3^2\left(\partial^2_i\sigma+3\tau\right)^2-2m_4^2\psi\left(\partial_i^2\sigma+3\tau\right)\Big]\,.
\end{align}
We can see that the fields $v$ and $\psi$ appear without time derivatives and are therefore auxiliary. We can integrate them out using their equations of motion,
\begin{align}
v&=\frac{1}{2m_1^2}\partial_0\tau\,;\label{eq:aux1}\\
\psi &= \frac{1}{2m_0^2}\left[2m_4^2(\partial_i^2\sigma + 3\tau)-\partial_i^2\tau\right]\,.
\label{eq:aux2}
\end{align}
Two special situations where this is not {possible} occur when $m_0=0$ or $m_1=0$. We will return to these later. Assuming $m_0,\;m_1\neq0$ for now, upon substitution of \cref{eq:aux1,eq:aux2} the scalar Lagrangian becomes
\begin{align}
\mathcal{L}_\mathrm{scalar} &= \frac{\mpl^2}{2}\bigg[
\frac32\tau\partial_0^2\tau - \frac{1}{2m_1^2}\partial_0^2\tau\partial_i^2\tau - \frac{1}{4m_0^2}(\partial_i^2\tau)^2 -\frac12\left(1-6\frac{m_4^2}{m_0^2}\right)\tau\partial_i^2\tau - 9 \frac{\mu^4}{m_0^2}\tau^2 \nonumber \\
&\hphantom{{}=\frac{\mpl^2}{2}\bigg[} -\frac{\mu^4}{m_0^2}(\partial_i^2\sigma)^2 + \frac{m_4^2}{m_0^2}\partial_i^2\tau\partial_i^2\sigma - 6\frac{\mu^4}{m_0^2}\tau\partial_i^2\sigma + \partial_0^2\tau\partial_i^2\sigma\bigg]\,.
\end{align}
The dispersion relations following from this are the solutions of
{
\begin{equation}
m_0^2\omega^4 + 2 \left(m_4^2-\frac{m_4^4-m_0^2m_3^2}{m_1^2}\right)p^2\omega^2 + 6\left(m_4^4-m_0^2m_3^2\right) \left(\omega^2 + \frac{p^2}{3}\right) + m_3^2p^4 = 0\,.
\label{eq:disp}
\end{equation}
}One can see that both scalar modes propagate in this case. Assuming that about the degravitating vacuum we have $m_i^2\sim\Lambda$, we can expand the solutions in the limit $p^2\ll\Lambda$ to find a massive mode with
\begin{equation}
\omega^2 = -6\frac{m_4^4-m_0^2m_3^2}{m_0^2} + \left[\frac13 - \frac{2}{m_0^2}\left(m_4^2 - \frac{m_4^4-m_0^2m_3^2}{m_1^2}\right)\right]p^2 + \oo\left(\frac{p^4}{\Lambda^2}\right)\,,
\end{equation}
and a massless mode with
\begin{equation}\label{eq:BDghostm2}
\omega^2=-\frac{p^2}{3}+\oo\left(\frac{p^4}{\Lambda^2}\right)\,.
\end{equation}

The massive mode {potentially} has both a tachyonic and a {gradient} instability, but {one can choose the values of $m_3^2$ and $m_0^2$ appropriately to avoid such instabilities.} 
On the other hand, the massless mode is debilitating because it has a gradient instability that cannot be removed by any choice of $m_i^2$. This is the mode that, as discussed above, could be removed in the decoupling limit by specific parameter tunings. We find that the mixing with gravity necessarily resurrects this mode regardless of parameter tunings.\footnote{Assuming, again, that $m_0$ and $m_1$ are non-zero, as otherwise the auxiliary structure changes.} This would not have been problematic had this mode been suppressed by the cutoff, because further corrections would be suppressed by the same amount, but the fact that it is completely unsuppressed means that it is resurrected by the mixing with gravity at all momenta. This is evident from the fact that the masses $m_i$ are completely absent from the dispersion relation~\eqref{eq:BDghostm2}. Of course, the mass term for the graviton must play some role in making this mode dynamical since it does not propagate in pure GR. Indeed, in GR the field $\psi$ vanishes identically from the action after varying with respect to $\psi$ to find that $\partial_i^2\tau=0$, leaving only a term $6\tau\partial_0^2\tau$, implying that $\omega^2=0$ and the mode does not propagate. 

\subsection{Curing pathologies by modifying the auxiliary structure}
\label{sec:m10}

Clearly the phase $m_2 = 0$ is {generically} unhealthy. {However,} as mentioned above, there are further parameter choices, or ``boundaries," that alter the auxiliary structure of the scalar action, namely $m_0=0$ and $m_1=0$. Since the auxiliary structure changes, the modes that remain and their stability properties can be dramatically different. Note that these are not necessarily fine-tunings or even tunings because it is possible to impose additional symmetries that enforce them.

The boundary $m_0=0$ removes one of the degrees of freedom, as is clear from the fact that it lowers the dispersion relation \eqref{eq:disp} from being quadratic in $\omega^2$ to linear.\footnote{{Technically, one should re-derive \cref{eq:disp} from the Lagrangian with $m_0=0$ to verify that it is unaltered. Varying the Lagrangian \eqref{eq:scalar1} (with $m_0=0$) with respect to $\psi$, one can eliminate $\sigma$ in terms of $\tau$ (\cref{eq:aux1} remains unchanged) to find a reduced Lagrangian in terms of $\tau$ and $\psi$. The dispersion relation that follows from this is precisely \cref{eq:disp} with $m_0=0$. }} The mode that remains is the unhealthy massless mode; taking $m_0\to0$ in \cref{eq:disp} and calculating the dispersion relation at low momenta, we find \cref{eq:BDghostm2} again. We note also that the tuning $m_0=0$ is not protected by any symmetry, and is therefore UV sensitive. {Hence we will not pursue this possibility any further.}

For the boundary $m_1=0$, however, the story changes significantly. In this case, \cref{eq:aux2} is unchanged, but the absence of the term $(\partial_iv)^2$ in \cref{eq:scalar1} means that $v$ is no longer auxiliary. It is instead a Lagrange multiplier that enforces the constraint
\begin{equation}\label{eq:vcons}
\partial_0\partial_i^2\tau=0\,,
\end{equation}
which implies either $\omega=0$ or $\tau=0$. The latter condition gives the reduced action~\cite{Dubovsky:2004sg}
\begin{equation} \label{eq:m1reduced}
\mathcal{L}_\mathrm{scalar}=\frac{\mpl^2}{2}\left(m_3^2-\frac{m_4^2}{m_0^2}\right)\left(\partial_i^2\sigma\right)^2\,,
\end{equation}
which implies the dispersion relation $p^2=0$ for $\sigma$. Thus, this boundary does not have any propagating degrees of freedom in the scalar sector and the instability is absent. We will therefore study the phase $m_1 = m_2 = 0$ in detail in the following section.

Finally, let us confirm that models with $m_1=m_2=0$ do not suffer from instantaneous modes. While the phase $m_1=0$ is known to contain instantaneous modes with dispersion relation $p^2=0$~\cite{Dubovsky:2004ud,Dubovsky:2005dw,Blas:2014ira}, we will demonstrate that these are removed in our case because we have further set $m_2=0$. Introducing a conserved matter energy-momentum tensor $T_\nm$, {which couples to the metric as usual as} 
\begin{equation}
\mathcal{L}_\mathrm{source} = - \frac{1}{2} h_{\mu\nu} T^{\mu\nu}\,,
\label{Lsource}
\end{equation}
then, for a {\it localized} energy-momentum source, the solution for the gauge-invariant scalar potentials $\Phi$ and $\Psi$ is~\cite{Dubovsky:2004ud,Dubovsky:2005dw}
\begin{equation}
\Phi=\Phi_\mathrm{GR}+\left(3-\frac{2m_0^2m_2^2}{m_4^4-m_0^2(m_3^2-m_2^2)}\right)\frac{m_2^2}{\partial_i^4}\frac{T_{00}}{2\mpl^2}\,; \qquad \Psi =\Psi_\mathrm{GR} \,,
\end{equation}
where $\Psi_\mathrm{GR}$ and $\Psi_\mathrm{GR}$ are the GR results. One can see that the correction to GR vanishes when one imposes our symmetry and sets $m_2=0$. The weak-field limit of the degravitating theory is therefore identical to that of GR, {at least for localized sources and to leading order in the derivative expansion.}\footnote{{Later on in \cref{sec:mixingwithgravity}, when we include higher-derivative corrections to stabilize the dispersion relation, we will find a small deviation from GR.}} Theories in which $m_1=m_2=0$ are therefore phenomenologically viable, and we will focus on these herein.

\subsection{Degravitation in the weak-field limit}
\label{weakfielddegravitation}

{The above conclusion that the weak-field limit agrees with GR pertained to localized sources. For a uniform source like the cosmological constant, the story is of course different. In this section we will show how degravitation works in the weak-field limit, similarly to the analysis of~\rcite{Dvali:2007kt}.} 

{This is most easily done in unitary gauge, where the Lagrangian is the sum of the quadratic Einstein-Hilbert action, the Lorentz-violating mass terms~\eqref{eq:LVMGaction} with $m_1 = m_2 = 0$, and a source term~\eqref{Lsource},
\begin{equation} 
\mathcal{L}_\textrm{weak-field} = \frac{\mpl^2}{2} \left[h^{\mu\nu}\mathcal{E}_{\mu\nu}^{\alpha\beta}h_{\alpha\beta} + m_0^2h_{00}^2+m_3^2h_{ii}h_{jj}-2m_4^2h_{00}h_{ii} \right]-\frac{1}{2}h_{\mu\nu} T^{\mu\nu}\,,
\label{Lweakfield}
\end{equation}
where $\mathcal{E}^\ab_\mn h_\ab$ is the linearized Einstein tensor, with $\mathcal{E}^\ab_\mn$ denoting the usual Lichnerowicz operator:
\begin{equation}
\mathcal{E}^\ab_\mn h_\ab = \frac14\left[\Box h_\mn -2\partial_\alpha\partial_{(\mu} h^\alpha_{\nu)} + \partial_\mu\partial_\nu h - \left(\Box h -\partial_\alpha\partial_\beta h^\ab\right)\eta_\mn\right]\,.
\end{equation}
With $T_{\mu\nu} = -\Lambda \Mp^2 \eta_{\mu\nu}$, the equations of motion are 
\begin{align}
\nonumber
 \mathcal{E}^\ab_{00} h_\ab + m_0^2 h_{00} - m_4^2 h_{ii} &= \frac{\Lambda}{2}\,;\\
  \mathcal{E}^\ab_{0i} h_\ab &= 0\,;\nonumber\\ 
 \mathcal{E}^\ab_{ij} h_\ab + \left(m_3^2 h_{kk} - m_4^2 h_{00}\right)\delta_{ij} &= -\frac{\Lambda}{2}\delta_{ij}\,.
\end{align}
These are solved by constant solutions:
\begin{align}\label{eq:DGWFL}
h_{00} &= \frac{1}{2} \frac{m_3^2-m_4^2}{m_0^2m_3^2 - m_4^4} \Lambda\,;\qquad h_{0i}=0\,;\qquad  h_{ij} = - \frac{1}{6} \frac{m_0^2-m_4^2}{m_0^2m_3^2 - m_4^4} \Lambda\delta_{ij} \,.
\end{align}
This is equivalent to Minkowski space-time because the metric $g_{\mu\nu} = \eta_{\mu\nu} + h_{\mu\nu}$ can be brought into the form $\operatorname{diag}(-1,1,1,1)$ by rescaling the coordinates.} {Even though we have assumed an Einstein-Hilbert kinetic term, the argument above is clearly oblivious to the kinetic structure. In particular, the degravitation mechanism will go through unscathed when we include higher-derivative corrections in \cref{UVinsensitivity}.} 

Note that the solution in equation \eqref{eq:DGWFL} is not unique. For instance, one can find locally (anti-)de Sitter solutions with
\begin{equation}
h_{00}=0\,;\qquad h_{0i}=\frac{\Lambda}{3}x_0x_i\,;\qquad h_{ij}=-\frac{\Lambda }{15}\vec{x}^2\delta_{ij}+\frac{\Lambda}{5}x_ix_j.
\end{equation}
This is the metric for linearized (anti-)de Sitter space written in the gauge where $h_{00}=h_{ii}=0$. One may reach this gauge from the more familiar slicing where
\begin{equation}
h_{00} = \frac{\Lambda}{3}\vec{x}^2\,;\qquad h_{0i}=0\,;\qquad h_{ij}=-\frac{\Lambda}{6}\vec{x}^2\delta_{ij}
\end{equation}
 by performing a linearized diffeomorphism with 
\begin{equation}
\xi^0=-\frac{\Lambda }{6}\vec{x}^2x^0\,;\qquad\xi^i=-\frac{\Lambda }{20}\vec{x}^2x^i.
\end{equation}

\section{Time-dependent volume-preserving spatial diffeomorphisms}
\label{sec:m10syms}

The analysis of the previous section has led our search for stable, phenomenologically-viable degravitating theories to those that have $m_2=0$ (to avoid a large tensor mass) and $m_1=0$ (to ensure stable perturbations). We further require that this choice be stable against quantum corrections, rather than being an artifact of a tuned Lagrangian. In other words, both of these parameter choices should be enforced by symmetries. 

\subsection{Candidate symmetries with \texorpdfstring{$m_1=0$}{m1=0}}

As discussed above, $m_2=0$ can only be enforced by time-independent volume-preserving spatial diffeomorphisms (V$_\mathrm{s}$Diff),
{
\begin{equation}
x^i\rightarrow x^i + \xi^i(x^j) \quad\textrm{with}\quad \partial_i\xi^i=0\,.
\end{equation}
Our first task is to identify the symmetries} that can be imposed in addition to V$_\mathrm{s}$Diff in order to set $m_1=0$.
We have found five such symmetries:
\begin{align}
x^i&\rightarrow x^i+\xi^i(t,x^j)\label{eq:ghostcondensation}\\
x^i&\rightarrow x^i + \xi^i(t)\label{eq:timedepspat}\\
t&\rightarrow t+\xi^0(t,x^i)\label{eq:weirdsymmetry}\\
t&\rightarrow t +\xi^0(t),\label{eq:UVsenssym} \\
x^i&\rightarrow x^i+\xi^i(t,x^j)\quad\textrm{with}\quad\partial_i\xi^i(t,x^j)=0\,.\label{eq:newsym}
\end{align}
The first four of these were identified in \rcite{Dubovsky:2004sg}, while symmetry \eqref{eq:newsym} is novel, to the best of our knowledge. It turns out that, of these five candidates, only the last one is consistent with our criteria for degravitation. Let us first tackle the unsuitable ones, in order:

\begin{enumerate}

\item {{\bf General time-dependent spatial diffeomorphisms,}} $x^i\rightarrow x^i+\xi^i(t,x^j)$: This is the symmetry group of a ghost condensate~\cite{ArkaniHamed:2003uy}. It is UV insensitive and fixes all the masses but $m_0$ to be zero. {As shown in \cref{sec:onescalar}, this theory fails} to degravitate, cf. criterion~\ref{item:finetune}.

\item {{\bf Time-dependent spatial translations,}} $x^i\rightarrow x^i + \xi^i(t)$: This is also UV insensitive but, unfortunately, it is incompatible with volume-preserving spatial diffeomorphisms. This is easy to see in terms of the St\"{u}ckelberg invariants. The invariants needed to impose symmetry~\eqref{eq:timedepspat} are $X$ and $W_n=\Tr(w)^n$ with $w^{ab}=B^{ab}-C^{0a} C^{0b}/X$, while the invariants needed for V$_\mathrm{s}$Diff are $X$, $Y$, and $b$. The overlapping subset is $X$, which is again a ghost condensate and is not suitable for degravitation.

\item {{\bf General space-dependent time reparametrizations,}} $t\rightarrow t+\xi^0(t,x^i)$: Models with this symmetry do not have a well-behaved Goldstone sector (decoupling limit)~\cite{Dubovsky:2004sg}. 

\item {{\bf Time reparametrizations,}} $t\rightarrow t +\xi^0(t)$: This symmetry is UV sensitive. The allowed higher-derivative operators reintroduce a vector mode with dispersion $\omega^2\propto p^2$ so that there is a massless mode that propagates at all energies but is not present in the low-energy effective theory. Moreover, there are two scalar modes with dispersion $\omega^4\propto p^2$, hence there is at least one ghost. Any action constructed with such a symmetry is therefore not a good low-energy EFT, cf. criterion~\ref{item:uvsens}.

\end{enumerate}

{The fifth symmetry, given in \cref{eq:newsym}, describes {time-dependent volume-preserving spatial diffeomorphisms}.  It enhances the time-independent V$_\mathrm{s}$Diff symmetry studied in \cref{sec:m20}, and sets $m_1= m_2 = 0$ while allowing for degravitation.} Written in terms of the St\"uckelberg fields $\Phi^A$, this is a symmetry acting on the spatial fields,\footnote{We write this symmetry non-linearly as we will be working with non-linear functions of $\Phi^A$ below.}
\begin{equation}
\Phi^a\to \Psi^a(\Phi^0,\Phi^b)\quad\textrm{with}\quad\det\left(\frac{\partial\Psi^a}{\partial\Phi^b}\right)=1\,.
\end{equation}
To our knowledge, this is the first time this symmetry has been studied in the context of Lorentz-violating massive gravity. The building blocks of this symmetry are the product $Yb$ and $X$.
{In other words, the general action in this case is
\begin{equation}
\frac{\mathcal{L}}{\sqrt{-g}} =\frac{\mpl^2}{2} m^2U(X,Yb)\,,
\label{eq:UXYbtheory}
\end{equation}
which is a special case of the finite-temperature superfluid action~\eqref{finiteTsuperfluid}. This theory} satisfies criteria~\ref{item:finetune},~\ref{item:GWmass}, and~\ref{item:stability}: it has all four scalars $\Phi^A$ present, the tensor mass vanishes due to the V$_\mathrm{s}$Diff symmetry, and, as shown in \cref{sec:m10}, it has no ghost, tachyonic, or gradient instabilities, as well as no instantaneous modes.

As an aside, if one constructs a theory from $Yb$ alone then the residual symmetry is the (Lorentz-invariant) transverse diffeomorphism (TDiff) symmetry,
\begin{equation}
\partial_\mu \xi^\mu=0\,. 
\end{equation}
Indeed, as shown by Padilla and Saltas~\cite{Padilla:2014yea} (see also \rcite{Kuchar:1991xd}), these theories result from applying the St\"{u}ckelberg procedure to TDiff theories such as unimodular gravity (after properly implementing the unimodular constraint using Lagrange multipliers). In fact, these theories are the St\"{u}ckelberged form of the Henneaux-Teitelboim action~\cite{Henneaux:1989zc}. In this sense, our theory is to TDiff gravity what the ghost condensate is to GR: the symmetry-breaking patterns are analogous.

\subsection{UV insensitivity of the \texorpdfstring{$X$}{X}, \texorpdfstring{$Yb$}{Yb} theory}
\label{UVinsensitivity}

Having arrived at the candidate symmetry~\eqref{eq:newsym}, we will now show that theories with this symmetry also satisfy criterion~\ref{item:uvsens}, {\it i.e.}, that they are UV insensitive.
{By UV insensitivity, we mean that} higher-order corrections to the Lagrangian do not reintroduce massless or low-energy modes not present in the low-energy theory. For simplicity we will analyze this in the decoupling limit, {\it i.e.}, neglecting the mixing with gravity, so we begin by summarizing this sector, which was studied by Dubovsky~\cite{Dubovsky:2004sg}.

The decoupling limit, or Goldstone sector, corresponds to performing a field redefinition that is \emph{pure gauge} on the action \eqref{eq:LVMGaction},
\begin{equation}
h_{\mu\nu}=\partial_\mu\pi_\nu+\partial_\nu\pi_\mu\,.
\end{equation}
One can further decompose the Goldstone fields $\pi_\mu$ into a transverse vector $\pi^\mathrm{T}_i$ and two scalars $\pi_0$ and $\pi_\mathrm{L}$, with the latter defined by
\begin{equation}\label{eq:newgoldstone}
\pi_i=\pi^\mathrm{T}_i+\frac{\partial_i\pi^\mathrm{L}}{\sqrt{-\partial_j^2}}\,.
\end{equation}
{The transverse vector $\pi^\mathrm{T}_i$ generates our symmetry, and hence should drop out of the action. Indeed, at the quadratic level the vector action is 
\begin{equation}\label{eq:gold_vector}
\mathcal{L}_\mathrm{vector}=\mpl^2\left[m_1^2\left(\dot{\pi}^\mathrm{T}_i\right)^2-m_2^2\left(\partial_i\pi^\mathrm{T}_j\right)^2\right]\,,
\end{equation}
which vanishes when $m_1=m_2=0$. The absence of $\pi^\mathrm{T}_i$ persists at the non-linear level, because time-dependent volume-preserving diffeormorphisms are a non-linear symmetry of the theory. We will come back to this point below.} 

Meanwhile, the action for the scalars $\pi_0$ and $\pi_\mathrm{L}$ (with $m_1=m_2=0$) is
\begin{equation}
\mathcal{L}_\mathrm{scalar}=\mpl^2\left[2m_0^2\dot{\pi}_0^2+ 4m_4^2\dot{\pi}_0\sqrt{-\partial_i^2}\pi_\mathrm{L}+2m_3^2\left(\partial_i\pi_\mathrm{L}\right)^2\right]\,.
\label{Lscalar}
\end{equation}
{The longitudinal mode $\pi_\mathrm{L}$ is auxiliary. We will verify below that it remains auxiliary beyond the decoupling limit, {\it i.e.}, when including mixing with gravity,
as well as when we include higher-derivative corrections compatible with the symmetries. We can integrate out $\pi_\mathrm{L}$ using its equation of motion,
\begin{equation}
\sqrt{-\partial_i^2}\pi_\mathrm{L} = -\frac{m_4^2}{m_3^2}\dot{\pi}_0\,.
\label{piLsolndecoupling}
\end{equation}
Substituting back into the scalar action~\eqref{Lscalar}, we obtain
\begin{equation}
\mathcal{L}_\mathrm{scalar} = 2\mpl^2\left(m_0^2- \frac{m_4^4}{m_3^2}\right) \dot{\pi}_0^2 \,.
\end{equation}
To ensure there is no ghost instability, the coefficient of the kinetic term should be positive:
\begin{equation}
m_0^2 - \frac{m_4^4}{m_3^2} > 0\,.
\label{eq:weakcond}
\end{equation}
Thus we have a single scalar mode with vanishing sound speed,
\begin{equation}\label{eq:goldstonedisp}
\left(m_0^2- \frac{m_4^4}{m_3^2}\right) \omega^2=0\,.
\end{equation}
Our analysis in~\cref{sec:m10} revealed that this mode is not resurrected by the mixing with gravity; nor, indeed, are any others.\footnote{{The field $\pi_L$ is auxiliary even non-linearly. The only metric component that contains time derivatives of $\pi_L$ is $h_{0i}$, and non-derivative interactions of this component are forbidden by the symmetry \eqref{eq:newsym}. Higher-order derivatives must always come with at least one additional spatial derivative so that the dispersion relation for this mode is $p^2=0$, indicating that it is auxiliary.  }} We will see below that this mode receives a small sound speed from higher-derivative corrections, and therefore describes a healthy scalar mode.} 

Generally speaking, we consider the theory to be UV sensitive if either the scalars, the vector, or both are reintroduced by higher-order operators without being suppressed by some cutoff.
Our strategy for determining UV sensitivity will be to write down all two-derivative operators compatible with the symmetry \eqref{eq:newsym} and then examine them using the methods of the previous section.\footnote{We ignore one-derivative terms by assuming parity invariance.} The allowed operators are (up to boundary terms)
\begin{align}\label{eq:operators}
(\partial_0 h_{00})^2\,;\quad(\partial_ih_{0i})^2\,;\quad(\partial_0h_{ii})^2\,;\quad \partial_0h_{00}\partial_ih_{0i}\,;\quad\partial_0h_{00}\partial_0h_{ii}\,;\quad\partial_ih_{0i}\partial_0h_{jj}\,; \nonumber\\
(\partial_ih_{00})^2\,;\quad (\partial_jh_{ii})^2\,;\quad \partial_i h_{00}\partial_ih_{jj}\,;\quad \partial_ih_{00}\partial_jh_{ij}\,;\quad \partial_ih_{kk}\partial_jh_{ij}\,,
\end{align}
where the last two are only invariant up to a total derivative, as well as
\begin{equation}
K_{ij}K^{ij}\,,
\label{eq:operators2}
\end{equation}
where $K_{ij}=\frac{1}{2}\left(\partial_0h_{ij}-\partial_ih_{0j}-\partial_jh_{0i}\right)$ is the extrinsic curvature on space-like hypersurfaces. We can also use the extrinsic curvature to construct the invariant $K^2$, which consists of terms already in the list \eqref{eq:operators}. Note that $K^2$ and $K_{ij}K^{ij}$ are invariant under full time- and space-dependent spatial diffeomorphisms ($x^i\rightarrow x^i+\xi^i(t,x^j)$), rather than just the volume-preserving version we are considering ($\partial_i \xi^i=0$). Note also that $\partial_0 h_{ij}$, which is not invariant under this symmetry on its own, can appear in the combinations $K^2$ and $K_{ij}K^{ij}$.

Recall that at lowest order in derivatives, the vector modes do not propagate. None of the two-derivative operators listed above contains the vectors, by virtue of their transverseness, so the vector modes are not reintroduced at the two-derivative level. Moreover, this is true at all orders in derivatives, for the following reason. The only metric components that contain the vector modes are $h_{0i}$ and $h_{ij}$. In order to construct $n$-derivative operators invariant under \cref{eq:newsym}, one must apply either a Kronecker delta or a partial derivative to each free spatial index (or construct an operator where this is the case after integration by parts). This process removes the vectors {entirely} since one has $\partial_i\pi_i^\mathrm{T}=0$. We therefore conclude that the vectors are absent non-linearly.

Another way to see the absence of additional degrees of freedom non-linearly is to consider the Hamiltonian for Lorentz-violating massive gravity in the Arnowitt-Deser-Misner (ADM) formalism. In terms of the ADM variables---the lapse $N$, the shift $N^i$, and the projected metric $\gamma_{ij}$---the metric is
\begin{equation}
g_{\mu\nu}\dd x^\mu\dd x^\nu=-N^2\dd t^2 + \gamma_{ij}\left(N^i\dd t + \dd x^i\right)\left(N^j\dd t + \dd x^j\right).
\end{equation}
The  Hamiltonian for Lorentz-violating massive gravity can be written in general as \cite{Comelli:2013txa}
\begin{equation}
H=\mpl^2\int \dd^3 \vec{x}\left[N^A\mathcal{H}_A + \pi^{ij}\gamma_{ij} + N\sqrt{\gamma} V(N,N^i,\gamma_{ij})\right],
\end{equation}
where $N^A=\{N,\,N^i\}$ act as Lagrange multipliers to impose the constraints $\mathcal{H}_A=\{\mathcal{H}_0,\,\mathcal{H}_A\}$ in GR, $\gamma=\det(\gamma_{ij})$, and $\pi^{ij}$ is the momentum conjugate to $\gamma_{ij}$. In the case of ghost condensation, the only non-derivative operator compatible with the symmetry is $h_{00}$ so that in general $V=V(N)$. For this reason, the shift vector still enforces three constraints whereas the lapse is now auxiliary. This is the reason only one extra degree of freedom propagates, even non-linearly. In our case, time-dependent volume-preserving spatial diffeomorphisms demand that $V=V(N,N\sqrt{\gamma})$.\footnote{This can easily be seen by working in unitary gauge, so that
\[X=-\frac{1}{N^2},\quad Yb=\frac{1}{\sqrt{-g}}=\frac{1}{N\sqrt{\gamma}}.\] The combination $N\sqrt{-g}$ is invariant under the full volume-preserving diffeomorphism symmetry, but the presence of $N$ in the potential breaks this down to time-dependent volume-preserving diffeomorphisms. In the case of unimodular gravity and other TDiff theories, Lorentz-invariance imposes that $V=V(N\sqrt{\gamma})$ only.} This implies that the shift is still a Lagrange multiplier and that our theory therefore propagates exactly three degrees of freedom non-linearly. The vector modes are therefore absent non-linearly and around any background. In the context of massive gravity, this implies that the Boulware-Deser ghost is absent in our theory.

{In the case of the scalars, the operators~\eqref{eq:operators} lead to a modification of the dispersion relation~\eqref{eq:goldstonedisp} of the form
\begin{equation}\label{eq:UV-insensitive!yay}
\left(m_0^2- \frac{m_4^4}{m_3^2}\right) \omega^2 = a\frac{p^4}{\mathcal{M}^2}+ b\frac{p^2\omega^2}{\mathcal{M}^2} + c\frac{\omega^4}{\mathcal{M}^2}\,,
\end{equation}
where $\mathcal{M}$ is some cutoff scale, naturally expected to be at most of order $\Lambda_2=\sqrt{m\mpl}$~\cite{ArkaniHamed:2002sp,Rubakov:2004eb}.
The constants $a$, $b$ and $c$ depend on the coefficients of the operators \eqref{eq:operators} and are therefore determined by the details of the UV physics.
The last two terms, proportional to $p^2\omega^2$ and $\omega^4$, can be neglected relative to the $\omega^2$ term, which was already present in the lowest-order action and is therefore unsuppressed by the cutoff.} 

The $p^4$ term, on the other hand, is crucial in stabilizing the dispersion relation. Of all the higher-derivative operators listed in \cref{eq:operators,eq:operators2},
the only operator that contributes to $p^4$ is
\begin{equation}
(\partial_i h_{0i})^2\,. 
\end{equation}
As a result of adding this operator, the scalar mode that had $\omega^2=0$ at low energies now propagates, with a Lorentz-violating dispersion relation $\omega^2\propto p^4/\mathcal{M}^2$. 
The presence or absence of a gradient instability for this mode depends on the sign of $a$, and so cannot be determined from the low-energy physics alone.\footnote{In this sense, the positivity of $a$ is a requirement imposed on any sensible UV completion of our model, much along the lines of other such positivity bounds in effective field theories \cite{Adams:2006sv}.} The absence of a ghost instability, however, can be determined from our knowledge of the low-energy EFT: it requires that we impose the weak condition~\eqref{eq:weakcond}. It is not difficult to find models that
satisfy this condition. Therefore, our theory is UV insensitive and stable.

Before moving on, it worth pausing to discuss the case where the residual symmetry is taken to be Lorentz-invariant transverse diffeomorphisms (TDiff), so that the only operator one can write down is $Yb$ (cf. the discussion above). In the unitary gauge, this implies that the action is given by the sum of the Einstein-Hilbert term and a function of $\sqrt{-g}$. This corresponds to taking $m_0=m_3=m_4$, as can be seen by expanding
\begin{equation}
\sqrt{-g}=1-\frac{1}{2}(h_{00}-h_{ii}) + \mathcal{O}(h^2),
\end{equation}
so that the massive part of the Lagrangian is
\begin{equation}
\mathcal{L}_{\rm mass}=\frac{\mpl^2m^2}{16}\left.\frac{\partial^2U(\sqrt{-g})}{\partial (\sqrt{-g})^2}\right|_{g=\eta}\left(h_{00}-h_{ii}\right)^2,
\end{equation}
which gives precisely $m_0=m_3=m_4$. Such actions typically arise in TDiff theories such as unimodular gravity \cite{Padilla:2014yea}. At leading order in derivatives there are no propagating scalar modes because the $\omega^2$ term on the left hand side of \cref{eq:UV-insensitive!yay} is absent. Higher-order derivative operators then reintroduce the scalar modes, which are required by Lorentz invariance to have a dispersion relation $\omega^2 = \pm p^2$. These theories are therefore UV sensitive. To our knowledge, this observation has not previously been made in the literature.

\subsection{{Mixing with gravity}}
\label{sec:mixingwithgravity}

The above analysis relied on the decoupling limit, and as such neglected the mixing with gravity. We now consider the effects of higher-derivative terms, including the mixing with gravity. To see why this is necessary, it is helpful to compare our degravitating theory with ghost condensation~\cite{ArkaniHamed:2003uy}, in which $U=U(X)$ and so only $m_0^2$ is nonzero. Both the ghost condensate and our theory have a zero-energy mode ($\omega^2=0$). At leading order this mode is fixed by the boundary conditions, which implies that we need to include the higher-order derivative operators identified in \cref{eq:operators,eq:operators2} in order to make the Goldstone fields dynamical and see a genuine modification of gravity.

As mentioned already, the operator that contains the higher-order gradients needed to give the Goldstones dynamics is $(\partial_i h_{0i})^2$. Hence it suffices to consider the quadratic action~\eqref{Lweakfield} supplemented by a $(\partial_i h_{0i})^2$ term,
\begin{equation}\label{eq:mixlag}
\mathcal{L} = \frac{\mpl^2}{2} \left[h^{\mu\nu}\mathcal{E}_{\mu\nu}^{\alpha\beta}h_{\alpha\beta} + m_0^2h_{00}^2+m_3^2h_{ii}h_{jj}-2m_4^2h_{00}h_{ii}\right] - 2\Mp^2\zeta (\partial_i h_{0i})^2-\frac{1}{2}h_{\mu\nu} T^{\mu\nu}\,,
\end{equation}
where $\zeta$ is a small, dimensionless parameter. {We will work} in Newtonian gauge. Since \cref{eq:mixlag} is not diffeomorphism-invariant this gauge fixing reintroduces the Goldstone fields {as follows:}
\begin{align}
h_{00}=2\Phi+2\dot{\pi}_0\,;\quad
h_{0i}=\partial_i\pi_0+\frac{\partial_i\dot{\pi}_\mathrm{L}}{\sqrt{-\partial_i^2}}\,;\quad\textrm{ and }\quad
h_{ij}= 2\Psi\delta_{ij}+2\frac{\partial_i\partial_j\pi_\mathrm{L}}{\sqrt{-\partial_i^2}}\,.
\end{align}
The action for the scalars is then 
\begin{align}
\mathcal{L}&=\frac{\mpl^2}{2} \left[4\Psi\partial_i^2\Phi-2\Psi\partial_i^2\Psi+4m_0^2\left(\Phi^2+2\Phi\dot{\pi}_0+\dot{\pi}_0^2\right)+4m_3^2\left(9\Psi^2-\pi_\mathrm{L}\partial_i^2\pi_\mathrm{L}-6\Psi\sqrt{-\partial_i^2}\pi_\mathrm{L}\right)\right.\nonumber\\&\left.-8m_4^2\left(3\Phi\Psi+3\Psi\dot{\pi}_0-\Phi\sqrt{-\partial_i^2}\pi_\mathrm{L}-\sqrt{-\partial_i^2}\pi_\mathrm{L}\dot{\pi}_0\right)-4\zeta\left(\partial_i^2\pi_0\right)^2-8\zeta\pi_\mathrm{L}\partial_i^2\sqrt{-\partial_i^2}\dot{\pi}_0\right]\nonumber\\& -\Phi T_{00} - \dot{\pi}_0 T_{00}\,,
\label{eq:LagrangianUV}
\end{align}
where we have taken the non-relativistic limit $\omega^2\ll p^2,\,T_{ii}\ll T_{00}$. The fields $\Psi$ and $\pi_\mathrm{L}$ can be eliminated via their equations of motion to find\footnote{The equations of motion are\[\partial _i^2(\Phi-\Psi)+6m_3^2\left(3\Psi-\sqrt{-\partial_i^2}\pi_\mathrm{L}\right)-6m_4^2(\Phi+\dot{\pi}_0)=0\,,\] and 
\[m_3^2\left(\partial_i^2\pi_\mathrm{L}+3\sqrt{-\partial_i^2}\Psi\right)-m_4^2\sqrt{-\partial_i^2}\left(\Phi+\dot{\pi}_0\right)+\zeta\partial_i^2\sqrt{-\partial_i^2}\dot{\pi}_0=0\,.\]}
\begin{align}
\Psi&=\Phi-6\zeta\dot\pi_0\label{eq:phipsi}\\
\sqrt{-\partial_i^2}\pi_\mathrm{L}&= 3\Phi-\frac{m_4^2}{m_3^2}\left(\Phi+\dot\pi_0\right)\,,
\label{eq:phipsi2}
\end{align}
where we have ignored sub-leading terms. As it should, \cref{eq:phipsi2} agrees with \cref{piLsolndecoupling} if we set $\Phi = 0$. 
Substituting these into the Lagrangian \eqref{eq:LagrangianUV} and ignoring the sources, we find, in Fourier space,
\begin{equation}\label{eq:UVmomlag}
\mathcal{L}=\frac{1}{2}\left(\pi_\mathrm{c}^\star~~\Phi_\mathrm{c}^\star\right)\begin{pmatrix} 
\omega^2-\zeta \frac{p^4}{\mu^2} & - \mathrm{i}\sqrt{2}\mu\omega \\
\mathrm{i}\sqrt{2}\mu\omega & -p^2+ 2\mu^2
\end{pmatrix}\begin{pmatrix} 
\pi_\mathrm{c}  \\
\Phi_ c  
\end{pmatrix}\,,
\end{equation}
where we have canonically normalized the fields,
\begin{equation}
\Phi=\frac{\Phi_\mathrm{c}}{\sqrt{2}\mpl}\,;\qquad \pi_0=\frac{\pi_\mathrm{c}}{2\mu\mpl}\,,
\end{equation}
and defined the mass scale
\begin{equation}
\mu^2= m_0^2-\frac{m_4^4}{m_3^2}\,,
\end{equation}
which is positive by virtue of the no-ghost condition~\eqref{eq:weakcond}.

The effects of mixing with gravity are found from \cref{eq:UVmomlag}, from which we can obtain the dispersion relation
\begin{equation}\label{eq:UVdisp}
\omega^2=-2\zeta p^2 +\zeta\frac{p^4}{\mu^2}\,.
\end{equation}
The mixing with gravity has introduced an instability at small momenta (large distances) that is stabilized by the presence of the $p^4$ term. This is precisely the same instability found in ghost condensate theories~\cite{ArkaniHamed:2003uy}.\footnote{The form of \cref{eq:UVmomlag} is exactly the same as the Fourier-space Lagrangian found in ghost condensate theories, which can be seen by identifying $\zeta=\alpha^2$ and {$\mu=M = \frac{m}{\sqrt{2}}$} (defined in \rcite{ArkaniHamed:2003uy}).} The maximum (imaginary) frequency for the instability is 
\begin{equation}\label{eq:omegains}
\omega_\mathrm{I}^2=-\zeta\mu^2\,,
\end{equation}
while the instability manifests on distances larger than $r_\mathrm{I}\sim \mu^{-1}$ (momenta $p \lsim \mu$). In particular, assuming a degravitating solution where $\mu^2\sim \Lambda\sim \rho_\mathrm{vac}/\mpl^2$, the instability manifests itself on distances larger than
\begin{equation}
r_\mathrm{I}\sim \left(\frac{\textrm{TeV}^4}{\rho_\mathrm{vac}}\right)^{\frac{1}{2}}\textrm{ mm}\,.
\end{equation}

At first sight, this appears to be debilitating for any useful theory because such an instability should be detectable in the laboratory or in space~\cite{Burrage:2017qrf,Sakstein:2017pqi}, but such a conclusion would be premature. Indeed, \cref{eq:omegains} indicates that it takes a time of order $t_\mathrm{I}\sim r_\mathrm{I}/\sqrt{\zeta}$ for the instability to arise, and it is possible that this timescale is longer than the age of the Universe. This is a consequence of Lorentz violation. Moreover, as first noted in \rcite{ArkaniHamed:2003uy}, retardation effects are important and must be accounted for before drawing any firm conclusions. 

Indeed, one can invert the matrix \eqref{eq:UVmomlag} to find the Green's function for the Newtonian potential,
\begin{equation}
G(\omega,p)=-\frac{1}{p^2}+\frac{2\zeta\mu^2}{\mu^2\omega^2 +4\zeta\mu^2p^2-\zeta p^4}\,.
\end{equation}
The second term represents the deviation from Newton's inverse-square law. Due to the harsh breaking of Lorentz invariance the precise nature of the instability depends strongly on whether or not the source is moving with respect to the rest frame of the superfluid. Indeed, sources in the rest frame produce oscillations in the Newtonian potential on distances shorter than $(r_\mathrm{I}/t_\mathrm{I})\times t$~\cite{ArkaniHamed:2003uy}, but sources moving with some velocity relative to the rest frame result in ``star tracks"~\cite{Dubovsky:2004qe} (see also \rcite{Peloso:2004ut}) where the instability only manifests in the wakes left behind by the object's motion. The probability of detecting these star tracks is negligible and is smaller for smaller values of $r_\mathrm{I}$~\cite{Dubovsky:2004qe}. In the case of ghost condensation, the likely situation is that the ghost condensate's rest frame coincides with the rest frame of the cosmic microwave background so that we are moving with a relative speed of $10^{-3}c$. One would therefore expect star tracks. In the case of a degravitating superfluid, the relevant situation is not so clear because one needs to construct a cosmological model before determining the rest frame of the superfluid. One might expect that we live in its rest frame since the effects of the cosmological constant are mitigated but this would imply that we live in Minkowski space, which is clearly not the case. In the absence of any concrete cosmological scenario, any bounds on $\zeta$ or statements about the viability of the theory would be premature. 

We can find a n\"{a}ive bound for $\zeta$ by considering the length scale at which $\pi_0$-mediated interactions become strongly coupled (which is not necessarily the cutoff for the effective field theory). The Einstein-Hilbert action contains interactions of the form $h^2\partial_i^2h$, where we have suppressed all indices. At low momenta the effective action is, schematically,
\begin{equation}
S=\int\dd^4 x\mpl^2\left[h^2\partial_i^2h-\mu^2h\left(\partial_0^2+c_s^2\partial_i^2\right)h\right]\,,
\end{equation}
where $c_s^2=2\zeta$ is the effective sound speed. Rescaling $x^i\rightarrow c_s x^i$ so that the quadratic part looks Lorentz-invariant, one has
\begin{equation}
S=\int\dd^4 x\mpl^2c_s\left[ h^2\partial_i^2h-\mu^2h\left(\partial_0^2+c_s^2\partial_i^2\right)\right]\,,
\end{equation}
so that the theory becomes strongly-coupled at a length scale $ r_\mathrm{c}=\Lambda_\mathrm{c}^{-1}\sim (\mpl^4\zeta)^{-1/4}$. Collisions at the LHC involve momentum transfers of order TeV so we would like the theory to be predictive up to this scale. Setting $\Lambda_\mathrm{c}>$ TeV the timescale for the instability satisfies
\begin{equation}
H_0 t_\mathrm{I}\le 10^{30} \frac{r_\mathrm{I}}{H_0^{-1}}=\mathcal{O}(10) \,,
\end{equation}
where we have assumed a TeV-scale vacuum energy so that $r_\mathrm{I}\sim$ mm. It is then possible to take the timescale for the instability to be longer than the age of the Universe although, again, Lorentz-violating effects mean that this is not necessary in order to have a phenomenologically-viable theory. 

Finally, we note that our theory differs from ghost condensation, {which has $\Phi = \Psi$, whereas our theory predicts} $\Psi=\Phi-6\zeta \dot{\pi}_0$,\footnote{Note that the term in \cref{eq:LagrangianUV} $\propto \zeta\pi_\mathrm{L}\partial_i^2\sqrt{-\partial_i^2}\dot{\pi}_0$ that gives rise to the deviation is absent if we only consider higher-derivative corrections arising from the extrinsic curvature operators \eqref{eq:operators2}. There is no symmetry argument for consider these terms solely and so we expect $\Psi\ne\Phi$ generically.} cf. \cref{eq:phipsi}. This implies that the parameterized post-Newtonian (PPN) parameter $\gamma\ne1$. Since $\pi_0$ is sourced by $\dot T_{00}$, such deviations would only show up for time-dependent sources. One might also expect that the superfluid could be detected by measuring the PPN parameters $\alpha_1$ and $\alpha_2$ since $\dot T_{00}=\partial_i T_{0i}=\rho \partial_i v_i$~\cite{Ip:2015qsa}. It would be interesting to study the effects of oscillating sources (such as the Sun) on the superfluid as well as the post-Newtonian predictions. We postpone such analysis until the uncertainty surrounding the rest frame of the superfluid is resolved.

\section{Degravitating solutions}
\label{sec:degrav}

Our analysis thus far has shown that one can construct theories that satisfy criteria~\ref{item:finetune}--\ref{item:uvsens} by taking $U=U(X,\,Yb)$, as we did in \cref{eq:UXYbtheory},
along with the condition~\eqref{eq:weakcond} so that no ghosts propagate in the UV. This theory can be seen as a result of imposing time-dependent volume-preserving spatial diffeomorphisms \eqref{eq:newsym}.

In this section we will show that such theories are able to successfully degravitate an arbitrary cosmological constant, cf. criterion~\ref{item:degrav}, and that they are phenomenologically viable. As a working example we will present two simple models that satisfy all of our criteria.

Consider the action~\eqref{eq:UXYbtheory} coupled to Einstein-Hilbert gravity, including a cosmological constant, 
\begin{equation}
S=\frac{\mpl^2}{2}\int\dd^4x\sqrt{-g}\left[R-2\Lambda+m^2U(X,Yb)\right]\,.
\end{equation}
{We remind the reader that the quantities $X$ and $Yb$ are defined in terms of the four phonon fields as}
\begin{align}
X &\equiv g^\mn\partial_\mu\Phi^0\partial_\nu\Phi^0\,; \\
Yb &\equiv -\frac{\varepsilon^{\mu\nu\rho\sigma}\varepsilon_{abc}}{6\sqrt{-g}}\partial_\mu\Phi^0\partial_\nu\Phi^a\partial_\rho\Phi^b\partial_\sigma\Phi^c = \frac{\det\partial\Phi}{\sqrt{-g}}\,.
\end{align}
This theory describes a superfluid at finite temperature~\cite{Nicolis:2011cs}.

We seek a degravitating solution, i.e., $g_\mn=\eta_\mn$, on the ground-state configuration
\begin{align}
\Phi^0 &= \alpha t\,; \nonumber \\
\Phi^a &= \beta x^a\,.
\end{align}
On this background, $X$ and $Yb$ are given by
\begin{align}
X &= -\alpha^2\,; \label{eq:XYb-1}\\
Yb &= \alpha\beta^3\,. \label{eq:XYb}
\end{align}
The energy-momentum tensor for $\Phi^A$ is
\begin{equation}\label{eq:UXYbTmn}
\frac{1}{\Mp^2}T_\mn = \frac{m^2}{2}\left[(U-Yb\,U_{Yb})g_\mn - 2U_X\partial_\mu\Phi^0\partial_\nu\Phi^0\right]\,.
\end{equation}
This takes the form of a perfect fluid, with energy density and pressure given by
\begin{align}
\rho &= \frac{m^2\Mp^2}{2}\left(2XU_X + Yb\,U_{Yb} - U\right)\,; \\
P &= \frac{m^2\Mp^2}{2}\left(U-Yb\,U_{Yb}\right)\,.
\end{align}
Note that, much like in the case of $P(X)$ theories, the null energy condition $\rho+P>0$ becomes $XU_X>0$.

In order to degravitate $\Lambda$, that is, to have a flat solution $g_\mn=\eta_\mn$, we need to have $T_\mn=\Mp^2\Lambda g_\mn$. As discussed above, we can split this into two requirements: $i)$ that $T_\mn$ be proportional to $\eta_\mn$; and $ii)$ that its magnitude be equal to $\Mp^2\Lambda$, in order to cancel the cosmological constant. We can see immediately (cf. the discussion in \cref{sec:onescalar}) that since there is no way to have $\partial_\mu\Phi^0\partial_\nu\Phi^0\propto\eta_\mn$, the first requirement sets
\begin{equation}\label{eq:DG1}
U_X = 0
\end{equation}
on the degravitating solution. One way to achieve this is to have $U$ depend on $Yb$ only, {in which case the theory would be invariant under transverse diffeomorphisms (TDiffs), such as unimodular gravity.  This is undesirable since, as argued earlier, TDiff theories are UV sensitive. }
Moreover, these theories cannot address the cosmological constant problem without fine-tuning~\cite{Padilla:2014yea}. {Instead,~\cref{eq:DG1} should be interpreted as being analogous to the ghost condensate condition.} It is a dynamical equation relating $\alpha$ and $\beta$ that may not be true away from the degravitating solution.

The second condition, $T_\nm=\Mp^2\Lambda g_\mn$, requires
\begin{equation}\label{eq:DG2}
U-\alpha\beta^3U_{Yb}=\frac{2\Lambda}{m^2}\,.
\end{equation}
\Cref{eq:DG1,eq:DG2} are two dynamical relations that yield solutions for $\alpha$ and $\beta$ given a specific choice of $U$.

Perturbing around the degravitating solution, the mass parameters $m_i^2$ are given by (see \cref{sec:masses})
\begin{align}\label{eq:mi}
m_0^2& = \frac18\alpha^2m^2\left(4\alpha^2U_{XX}-4\alpha\beta^3U_{X\,Yb}+\beta^6U_{Yb\,Yb}\right)\,; \nonumber \\
m_1^2&= m_2^2=0\,;\nonumber \\
m_3^2 & = \frac18\alpha^2\beta^6m^2U_{Yb\,Yb}\,;\nonumber \\
m_4^2 & = \frac18\alpha^2\beta^3m^2\left(\beta^3U_{Yb\,Yb}-2\alpha U_{X\,Yb}\right)\,.
\end{align}
One can see that, as expected, $m_1=m_2=0$, which implies ghost freedom, as shown in \cref{sec:m10}.
The stability condition~\eqref{eq:weakcond} reduces to 
\begin{equation}
\frac{1}{U_{Yb\,Yb}}\left(U_{Yb\,Yb} U_{XX} - U_{X\,Yb}^2\right) > 0\,.
\end{equation}
%}

\subsection{Toy examples}
\label{sec:toyexample}

{We now provide two working examples of $U(X,Yb)$ that satisfy all of our criteria. The first example consists of a ghost condensate action plus a quadratic term in $Yb$: 
\begin{equation}\label{eq:specmodel1}
U(X,Yb)= \frac{K_1}{2} \left(X+1\right)^2 +  \frac{K_2}{2} (Yb)^2\,,
\end{equation}
where $K_1$ and $K_2$ are constants. Crucially, they will not be taken to depend in any way on $\Lambda$, as this would {be tantamount to fine-tuning}.
Note that we have not included a term proportional to $Yb$ since $\sqrt{-g}Yb = \det(\partial_\mu\Phi^A)$ is independent of the metric and therefore does not contribute to the metric equations of motion. Thus $(Yb)^2$ is the lowest-order non-trivial term in a power series expansion.} 

Using \cref{eq:XYb-1,eq:XYb}, the degravitation conditions~\eqref{eq:DG1} and~\eqref{eq:DG2} are satisfied by
\begin{equation}\label{eq:degravonshell1}
\alpha =1\,; \qquad \beta^6 = - \frac{4\Lambda}{K_2m^2}\,.
\end{equation}
It follows that we need $K_2 < 0$ to degravitate $\Lambda > 0$, and vice versa for $\Lambda < 0$. The masses are
\begin{align}
m_0^2&=\frac{1}{2}\left(K_1m^2-\Lambda\right);\\
m_1^2&= m_2^2=0\,;\nonumber \\
m_3^2&=m_4^2=-\frac{\Lambda}{2}.
\end{align}
Meanwhile, the stability condition~\eqref{eq:weakcond} requires
\begin{equation}
K_1  > 0.
\end{equation}
Therefore, we can degravitate an arbitrary cosmological constant provided that we choose the signs of $K_1$ and $K_2$ correctly.

Another simple choice for $U$ that successfully degravitates is
\begin{equation}\label{eq:specmodel}
U(X,Yb)=-X +\gamma X\,Yb-\frac{\lambda}{2}(Yb)^2\,,
\end{equation}
where $\lambda$ and $\gamma$ are arbitrary dimensionless constants, once again independent of $\Lambda$. The degravitation conditions~\eqref{eq:DG1} and~\eqref{eq:DG2} in this case imply
\begin{align}\label{eq:degravonshell}
\gamma\alpha\beta^3&=1\,; \nonumber\\
\alpha^2\left(1+\frac12\lambda\beta^6\right)&=\frac{2\Lambda}{m^2}\,.
\end{align}
These determine the integration constants to be
\begin{align}\label{eq:alphabeta}
\alpha&=\sqrt{\frac{2\Lambda}{m^2}-\frac{\lambda}{2\gamma^2}}\,;\nonumber \\
\beta&=\left(\frac{2\gamma^2\Lambda}{m^2}-\frac{\lambda}{2}\right)^{-\frac{1}{6}}\,,
\end{align}
where we have assumed $4\gamma^2\Lambda>\lambda m^2$ in order for this solution to exist. This model can therefore degravitate a positive cosmological constant with $\Lambda>\lambda m^2/4\gamma^2$. The masses are given by
\begin{align}
m_0^2& = \frac{\lambda}{8\gamma^2}m^2-\Lambda\,; \nonumber \\
m_1^2&= m_2^2=0\,;\nonumber \\
m_3^2 & = -\frac{\lambda}{8\gamma^2}m^2\,;\nonumber \\
m_4^2 & = -\frac\Lambda2\,.
\label{eq:mspecc}
\end{align}
It is straightforward to verify that the condition~\eqref{eq:weakcond} for the absence of ghosts at high energies is satisfied,
\begin{equation}
m_0^2-\frac{m_4^4}{m_3^2} = \frac{\lambda m^2}{8\gamma^2}\left(1-\frac{4\gamma^2\Lambda}{\lambda m^2}\right)^2 > 0\,,
\end{equation}
so that the theory is a good low-energy EFT provided we take $\lambda>0$.

\section{Conclusions}
\label{sec:conc}

In this paper we have presented a novel solution to the cosmological constant problem. In our model, a cosmological finite-temperature superfluid pervades the Universe and cancels out the cosmological constant dynamically when its internal degrees of freedom, a quartet of scalar fields describing the superfluid's internal coordinate space, spontaneously acquire Lorentz-violating vacuum expectation values. This cancellation is automatic and is a result of the equations of motion so that no fine-tuning is necessary. In unitary gauge, this theory admits an interpretation as a Lorentz-violating mass term for the graviton.

The Lorentz violation allowed us to evade Weinberg's no-go theorem and the recent theorem of \rcite{Niedermann:2017cel}, which assume Poincar\'{e} invariance and Lorentz invariance, respectively. The usual problem of not being able to simultaneously use the Vainshtein mechanism to screen fifth forces and degravitate a large cosmological constant in massive gravity is avoided since the vDVZ discontinuity is absent in Lorentz-violating theories.

We have scanned the various residual symmetries that could be retained after diffeomorphism breaking in an attempt to find phases that have no massive tensor modes so that bounds on the graviton mass are avoided. (The graviton mass-squared on the degravitating solution is generically of order the degravitated cosmological constant.) This led us to consider the phase of massive gravity where $m_2^2=0$ (imposed by time-independent volume-preserving spatial diffeomorphisms), but this was not sufficient because the spectrum contained a mode with dispersion relation $\omega^2=-p^2/3$. We showed that this can be eliminated by setting $m_1^2=0$ and examined candidate theories that can enforce this. We identified time-dependent volume-preserving spatial diffeomorphisms of the superfluid as the unique symmetry that enforces $m_1^2=m_2^2=0$ and gives rise to a UV insensitive theory. This corresponds to transverse spatial diffeomorphisms $\partial_i\xi^i(t,x^j)=0$ and, to our knowledge, this is the first time it has been identified and studied in the context of Lorentz-violating massive gravity.

The symmetry removes the propagating scalar modes so that the scalar dispersion relation is $\omega^2=0$, but this can be lifted by higher-derivative operators. We showed that our symmetry is UV insensitive in the sense that higher-order operators parameterized by a dimensionless number $\zeta$ give rise to the Lorentz-violating dispersion relation $\omega^2\propto p^4/\mathcal{M}^2$ so that these modes are suppressed below the cutoff $\mathcal{M}=(m_0^2-m_4^4/m_3^2)/\zeta$. There is a mild constraint, $m_0^2-m_4^4/m_3^2>0$, that must be imposed to ensure that the scalar is not a ghost. Mixing with gravity induces a Jeans-like instability ($\omega^2=-2\zeta p^2+\zeta p^4/\mathcal{M}^2$), but this is not problematic because Lorentz-violating effects ensure that the timescale for this to appear is long at large distances ($\gsim\mathcal{O}(\mathrm{mm})$ for degravitation of a TeV-scale cosmological constant). We argued that a TeV-scale cosmological constant ($\Lambda\mpl^2\sim \mathcal{O}(\textrm{TeV}^4)$) can be degravitated without the theory becoming strongly coupled and that this could be pushed to higher energies if need be.

We discussed the phenomenology of the weak-field limit and conjectured that oscillating objects (or, to be more, precise, time-dependent densities) would source differences between the gravitational potentials $\Phi$ and $\Psi$, and that Lorentz-violating effects may be present in the solar system. A specific cosmological model is needed in order to make quantitative predictions since Lorentz-violating effects mean that the relevant situation depends strongly on our motion relative to the superfluid's rest frame. For this reason, we postpone a detailed study for followup work.

Finally, we studied how degravitation works in our theory using two specific models to illustrate the pertinent features. We showed how this works in general, and in the weak-field limit, which is an important check that the large cosmological constant in our action does not manifest in the solar system and modify its dynamics. A full cosmological analysis of our model will appear in a forthcoming publication.

In summary, we have presented a new solution to the cosmological constant problem that is both UV insensitive and makes novel phenomenological predictions. Along the way we have shown that the phase of Lorentz-violating massive gravity with $m_2^2=0$ is pathological unless one also has $m_1^2=0$. We identified a new symmetry ($\partial_i\xi^i(t,x^j)=0$) that enforces this and have shown that our theory is a consistent infrared modification of gravity. 

\acknowledgments

We are extremely grateful for conversations with {Paolo Creminelli, Andrei Khmelnitsky, Tony Padilla, Luigi Pilo, and especially Sergei Dubovsky}. We would like to thank Marco Crisostomi, Kurt Hinterbichler, Kazuya Koyama, Riccardo Penco, and Glenn Starkman for enlightening discussions. {JS and ARS are supported by funds provided to the Center for Particle Cosmology by the University of Pennsylvania. JK is supported in part by NSF CAREER Award PHY-1145525, US Department of Energy (HEP) Award DE-SC0017804, NASA ATP grant NNX11AI95G, the Charles E. Kaufman Foundation of the Pittsburgh Foundation, and a W. M. Keck Foundation Science and Engineering Grant.}

\appendix

\section{Calculation of the graviton mass parameters}
\label{sec:masses}

In this Appendix we present some details of the calculation of the graviton mass parameters $m_i^2$ for the healthy degravitating theory $U(X,Yb)$ which we have arrived at in the main body of this paper. It is straightforward to generalize the procedure in this section to other models.

Consider the Lagrangian for $\Phi^A$,
\begin{equation}
\mathcal{L}_\mathrm{mass}=\frac{\mpl^2}{2}\sqrt{-g}\left(-2\Lambda+ m^2U(X,Yb)\right)\,.
\end{equation}
We will consider linear fluctuations about flat solutions in unitary gauge,
\begin{align}
g_\mn &= \eta_\mn + h_\mn\,; \nonumber \\
\Phi^0&=\alpha t\,; \nonumber \\
\Phi^a&=\beta x^a\,. \label{eq:uniperts}
\end{align}
In this case, the linearized action is of the form \eqref{eq:LVMGaction}~\cite{Rubakov:2004eb,Dubovsky:2004sg}
\begin{equation}
\mathcal{L}_\mathrm{mass}=\frac{\mpl^2 }{2}\left[m_0^2h_{00}h_{00}+2m_1^2h\oi h\oi-m_2^2h\iij h\iij+m_3^2h\ii h\jj-2m_4^2h\ooo h\ii\right]\,.
\end{equation}
The goal of this Appendix is to relate the $m_i^2$ parameters to $U(X,Yb)$ and its derivatives evaluated on a degravitating flat background.

We start by calculating the scalar operators
\begin{align}
X &= g^\mn\partial_\mu\Phi^0\partial_\nu\Phi^0\,; \\
Y &= -\frac{\varepsilon^{\mu\nu\rho\sigma}\varepsilon_{abc}}{6\sqrt{-g} b}\partial_\mu\Phi^0\partial_\nu\Phi^a\partial_\rho\Phi^b\partial_\sigma\Phi^c\,; \\
b &= \det(g^\mn\partial_\mu\Phi^a\partial_\nu\Phi^b)\,,
\end{align}
as well as the metric determinant $\sqrt{-g}$. In unitary gauge, we can write the scalar operators as
\begin{align}
X &= \alpha^2g^{00}\,; \\
Y &= \frac{\alpha\beta^3}{\sqrt{-g}b}\,; \\
b &= \det(g^{ij})\,.
\end{align}
Note that in calculating $Y$ it is helpful to replace the cumbersome contractions of Levi-Civita symbols with a determinant of $\partial_\mu\Phi^A$, which can easily be computed since in unitary gauge $\partial_\mu\Phi^A=\operatorname{diag}(\alpha,\beta\delta^a_i)$ is diagonal, {\it i.e.},
\begin{align}
-\frac16\varepsilon^{\mu\nu\rho\sigma}\varepsilon_{abc}\partial_\mu\Phi^0\partial_\nu\Phi^a\partial_\rho\Phi^b\partial_\sigma\Phi^c &= -\frac1{4!}\varepsilon^{\mu\nu\rho\sigma}\varepsilon_{ABCD}\partial_\mu\Phi^A\partial_\nu\Phi^B\partial_\rho\Phi^C\partial_\sigma\Phi^D\nonumber\\
&= \det(\partial_\mu\Phi^A) \nonumber \\
&= \alpha\beta^3\,.
\end{align}
Finally, recall that we are particularly interested in models where the Lagrangian only depends on $Y$ and $b$ through their product,
\begin{align}
Yb &= \frac{\det(\partial_\mu\Phi^A)}{\sqrt{-g}} \nonumber\\
&= \frac{\alpha\beta^3}{\sqrt{-g}}\,.
\end{align}
It is now straightforward to expand $X$, $Yb$, and $\sqrt{-g}$ to quadratic order in $h_\mn$, finding
\begin{align}
X&=-\alpha^2\left(1+h\ooo+h\ooo h\ooo-h\oi h\oi\right)\,;\\
Yb &= \alpha\beta^3\left[1+\frac{1}{2}(h\ooo-h\ii) +\frac{3}{8}h\ooo h\ooo - \frac14h\ooo h\ii + \frac{1}{8}\left(h\ii h\ii-4h\oi h\oi +2h\iij h\iij\right)\right]\,; \\
\sqrt{-g} &= 1 + \frac12\left(h_{ii}-h_{00}\right) + \frac18\left(h_{ii}h_{ii} - h_{00}h_{00} - 2h_{00}h_{ii} + 4h_{0i}h_{0i} - 2h_{ij}h_{ij}\right)\,.
\end{align}

The easiest method to compute the $m_i^2$ parameters is to take two partial derivatives of $\mathcal{L}$ with respect to $h_\mn$ and evaluate these on $h_\mn=0$, {\it i.e.},
\begin{align}
m_0^2 &= \frac{1}{\Mp^2}\left.\frac{\partial^2\mathcal{L}_\mathrm{mass}}{\partial h_{00}\partial h_{00}}\right|_{h_\mn=0}\,; \\
m_1^2 &= \frac{1}{2\Mp^2}\left.\frac{\partial^2\mathcal{L}_\mathrm{mass}}{\partial h_{0i}\partial h_{0i}}\right|_{h_\mn=0}\,; \\
m_2^2 &= -\frac{1}{\Mp^2}\left.\frac{\partial^2\mathcal{L}_\mathrm{mass}}{\partial h_{ij}\partial h_{ij}}\right|_{h_\mn=0}\,; \\
m_3^2 &= \frac{1}{\Mp^2}\left.\frac{\partial^2\mathcal{L}_\mathrm{mass}}{\partial h_{ii}\partial h_{jj}}\right|_{h_\mn=0}\,; \\
m_4^2 &= -\frac{1}{\Mp^2}\left.\frac{\partial^2\mathcal{L}_\mathrm{mass}}{\partial h_{00}\partial h_{ii}}\right|_{h_\mn=0}\,.
\end{align}
We will illustrate this concretely for $m_1^2$ and demonstrate this it vanishes identically for any $U(X,Yb)$. Taking a single derivative of $\mathcal{L}_\mathrm{mass}$ with respect to $h_{0i}$, we find
\begin{equation}
\frac{\partial\mathcal{L}_\mathrm{mass}}{\partial h_{0i}} = \frac{\Mp^2}{2}\left(m^2U-2\Lambda + 2m^2\sqrt{-g}\alpha^2U_X-m^2\sqrt{-g}\alpha\beta^3U_{Yb}\right)h_{0i}\,.
\end{equation}
Note that, because we are going to take another $h_{0i}$ derivative and then set $h_\mn\to0$ at the end, we can simply consider the terms in parentheses to be evaluated on the background. Taking another derivative with respect to $h_{0i}$, we find
\begin{equation}
\frac{\partial^2\mathcal{L}_\mathrm{mass}}{\partial h_{0i}\partial h_{0i}} = \frac{\Mp^2}{2}\left(m^2U-2\Lambda + 2m^2\alpha^2U_X-m^2\alpha\beta^3U_{Yb}\right)\,.
\end{equation}
Recall, however, that the degravitation conditions for this model (\cref{eq:DG1,eq:DG2}), which are nothing other than the equations of motion for $\Phi^A$, are
\begin{align}
U_X &= 0\,; \\
m^2U-2\Lambda &= m^2\alpha\beta^2U_{Yb}\,,
\end{align}
when $U$, $U_X$, and $U_{Yb}$ are evaluated on the background. So we see that $m_1^2=0$ exactly, as promised. Calculating the other $m_i^2$ and imposing the background equations of motion in the same way, we obtain \cref{eq:mi}.

\bibliographystyle{jhep}
\bibliography{ref}

\end{document}